\documentclass[a4paper,10pt]{article}

\usepackage{graphicx}
\usepackage{color}

\newtheorem{fig}{Figure}

\def\leurre{\noindent\leftskip0pt\small\baselineskip 10pt}

\def\encadre#1#2{%
\setbox100=\hbox{\kern#1{#2}\kern#1}
\dimen100=\ht100 \advance \dimen100 by #1
\dimen101=\dp100 \advance \dimen101 by #1
\setbox100=\hbox{\vrule height \dimen100 depth \dimen101\box100\vrule}
\setbox100=\vbox{\hrule\box100\hrule}
\advance \dimen100 by .4pt \ht100=\dimen100
\advance \dimen101 by .4pt \dp100=\dimen101
\box100
\relax
}

\def\ligne#1{\hbox to \hsize{#1}}
\def\PlacerEn#1 #2 #3 {\rlap{\kern#1\raise#2\hbox{#3}}}

\font\rmx=cmr10
\font\rmxii=cmr12

\font\itx=cmti10
\font\ttx=cmtt10

\title{A proposal for a Chinese keyboard for cellphones, smartphones, ipads and tablets
\vskip 15pt
\rmxii
\ligne{\hfill Maurice {\sc Margenstern}$^1$,\hskip 30pt Lan {\sc Wu}$^2$\hfill} 
\vskip 15pt
\rmx\baselineskip=12pt
\ligne{\hfill
{\small$^1$}\vtop{\parindent 0pt\leftskip 0pt\hsize=150pt
\ligne{Universit\'e de Lorraine,\hfill}
\ligne{LITA, EA 3097,\hfill}
\ligne{Campus du Saulcy,\hfill}
\ligne{57045 Metz Cedex, France,\hfill}
}
\hskip 50pt
{\small$^2$}\vtop{\parindent 0pt\leftskip 0pt\hsize=90pt
\ligne{in private capacity\hfill}
\hfill
}
}
\ligne{{\itx email:} {\ttx maurice.margenstern@univ-lorraine.fr}
\hfill}
}

\begin{document}
\maketitle

\vskip 10pt
\begin{abstract}
In this paper, we investigate the possibility to use two tilings of the hyperbolic plane
as basic frame for devising a way to input texts in Chinese characters into messages 
of cellphones, smartphones, ipads and tablets. 
\end{abstract}
\vskip 10pt

\def\cqfd{\hbox{\kern 2pt\vrule height 6pt depth 2pt width 8pt\kern 1pt}}
\def\Hii{$I\!\!H^2$}
\def\Hiii{$I\!\!H^3$}
\def\Hiv{$I\!\!H^4$}
\def\norm{\hbox{$\vert\vert$}}
\section{Introduction}

   In a previous paper, \cite{acri}, the first author presented a proposal for a
Japanese keyboard for cellphones. He presented his ideas first at
AUTOMATA'2005{} in Gdansk, {\sc Poland}. A few Japanese colleagues were there present.
They highly appreciated the project and decided to join it. Later on, we call the project
described in~\cite{acri} as the {\it Japanese project}.

   Here, the paper deals with Chinese language, and we shall later reference the project as 
the {\it Chinese project}. It is important to indicate at this point that we illustrate
our example using the simplified Chinese characters. It is not difficult to adapt the
same method with traditional characters. 

   There is a deep difference between both projects. In the Japanese
project, the question of kanji's was simply evoked not even partially implemented. The
writing of the texts was based on hiragana's and katakana's, Japanese syllabic alphabets
which allow anyone to phonetically write down texts of the Japanese language. In the Chinese
project, we address the question of writing texts in Chinese characters, and the experimental
aspect, although at a very tiny toy example, actually yields a text in Chinese characters.

   In Section~\ref{main}, we describe the main features of the Chinese project.
project. As in the Japanese project, the Chinese one makes use of tessellations in the hyperbolic
plane. In the Japanese project, we used one tessellation, the pentagrid, see 
Subsection~\ref{hyptilings} while in the Chinese project, we use two of them, the pentagrid
and the heptagrid, see Subsection~\ref{hyptilings} and Figure~\ref{tilings}.
In Section~\ref{implement}, we describe our implementation and how we proceeded to perform
the toy example illustrating the paper.

\section{Main ideas}
\label{main}

    The main ideas consists in merging a part of the today technique used in typing 
Chinese texts on a computer with the technique used in the Japanese project.

    As an example of the today technique used for typing Chinese texts on a computer,
we can take the algorithm used by Google. Once you selected the Chinese keyboard,
you are given the possibility to use a font of simplified characters or a font of
traditional characters. Then you type the pin-yin of each character. In a more or less 
predictive way, the system opens a window where you are given the choice for the character
you wish. If you type the pin-yin of a single character, the system gives you a numbered list
of five characters you can select with the mouse and then clicking on the expected character.
In fact there are usually more than five characters for a given syllab. This is indicated by an
arrow. If the wished character is not in the list, clicking on the arrow displays a new list
and this can be repeated until you find the character or all characters corresponding
to that pin-yin have been displayed.  
     
    Also note that the just described system does not allow you to mention the tone of
the syllab.

    In our Chinese project, we also use first a pin-yin approach. Now, this time we allow
the user to type an optional information on the tone: 1, 2 3 or~4 depending on the number of
the considered tone. It is known that the tones of the standard mandarin Chinese are ordered
from~1 to~4. The lack of tone is denoted by~0. Now, we introduce a difference. As on a cell 
phone there is no room for an alphabetic typing, we use a tiling of the hyperbolic plane
to diplay the letters needed by the user. We split the process into two parts as,
according to the phonetic properties of syllabs in the Chinese language, we may separate
a syllab into first, a consonent, and then a vowel. We use another tiling of the hyperbolic
plane in order to display the characters associated to the syllab selected by the user.
In Section~\ref{implement}, we precisely describe the method which is also illustrated on 
an example. 

\section{Pentagrid and heptagrid}
\label{pentahepta}

    In this section, we sketchily remind what should be known about hyperbolic geometry
in Subsection~\ref{hypgeom}. Then, in Subsection~\ref{hyptilings} we focus on tilings of
the hyperbolic plane, especially on the two ones we shall use in our proposal: the pentagrid
and the heptagrid.

\subsection{Hyperbolic geometry}
\label{hypgeom}

Hyperbolic geometry was found after a two-thousand year research about an axiom
of Euclid's {\it Elements}. The axiom says that in the plane, through a point~$P$ out of
a straight line~$\ell$ there is exactly one straight line~$p$ which passes through~$P$ and
which is parallel to~$\ell$, meaning that it does not cut~$\ell$. For a reason we cannot explain
here, people were convinced that this property could be proved from the other axioms of
the {\it Elements}. It took around two thousand years to realize that Euclid was right:
his axiom cannot be proved from the others. But the answer was a source of deep astonishment.
It was not a simple answer: it was at the same time, the discovery of a new world, as coherent
as the one to which people were used. Less than a hundred year after the discovery, it turned
out that hyperbolic geometry was used in the foundations of an important theory of physics,the 
theory of relativity.

   Let us just mentioned here that a model of this geometry was discovered around 1870
and that the most presently used modelis of this geometry, so called Poincar\'e's models
were found in 1882. In this paper, we make use of Poincar\'e's disc model, illustrated 
by Figure~\ref{poincare}.

\vtop{
\vspace{-620pt}
\ligne{\hfill\includegraphics[scale=1]{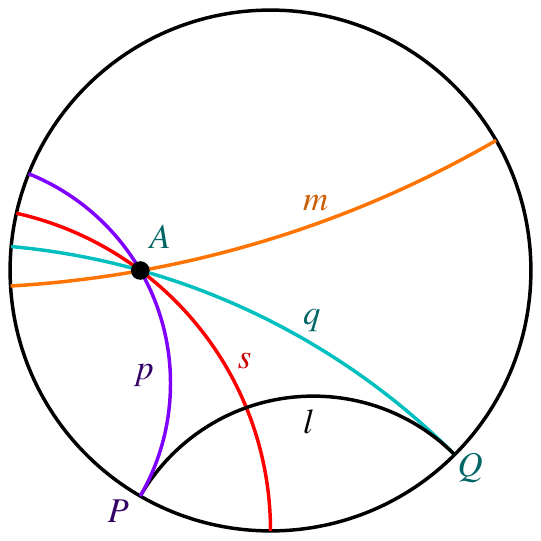}
\hfill}
\vspace{-5pt}
\begin{fig}
\label{poincare}
\leurre
The Poincar\'e's disc.
\end{fig}
}

   In the Figure, we can see a circle~$C$ which we call the set of {\bf points at infinity}.
The points which lie inside~$C$ constitute the points of the hyperbolic plane.
Note that the points at infinity do not belong to the hyperbolic plane. A line of the
hyperbolic plane is represented by the trace, inside~$C$, of either a diameter of~$C$
or of a circle which is orthoogonal to~$C$. Figure~\ref{poincare} represents several lines
of the hyperbolic plane. We can see that, in the figure, $\ell$ does not pass through~$A$
which is a point of the hyperbolic plane. Now, the line~$s$, which passes through~$A$,
cuts~$\ell$ in the hyperbolic plane. The figure shows that there are two lines, $p$ and~$q$, 
passing through~$A$ which have a common point with~$\ell$ which is a point at infinity: and 
so the lines~$p$ and~$q$ are called {\bf parallel} to~$\ell$. Now, there is another kind
of line, represented by~$m$ in Figure~\ref{poincare}, which passes through~$A$ but does not
meet~$\ell$ neither inside~$C$, nor on~$C$ and nor outside~$C$. Such a line is called 
{\bf non-secant} with~$\ell$.

   We have no room here to go further about the properties of hyperbolic geometry where
triangles have new properties and were new objects also have many fascinating properties.
The interested reader is referred to~\cite{mmbook1,mmbook2,mmbook3}. In~\cite{mmbook2},
an account of the Japanese project can be found. In~\cite{mmbook3}, there is a short 
historical account of the discovery of hyperbolic geometry. In~\cite{mmbook1,mmbook3} there is
an introduction to hyperbolic geometry.

\subsection{Tessellations in the hyperbolic plane}
\label{hyptilings}

   In 1882, Henri Poincar\'e proved a very important property of the hyperbolic plane, namely
the existence of infinitely many tilings generated by the following process: we start with
a convex regular polygon, called the {\bf basis} and we replicate it by reflection in 
its sides and, recursively we replicate the images in reflection in their sides. 
These images together with the basis itself are called {\bf copies} of the basis. When 
the copies cover the plane in such a way that pairwise copies never intersect their interiors, 
we say that we have a tiling. A tiling obtained by the just described process is calle a 
tessellation.

    If we denote by~$p$ the number of sides of the basis and by~$q$ the number of copies we can
put, sharing a common vertex~$V$, not overlapping and covering a neighbourhoodof~$V$,
Poincar\'e's theorem that we have a tessellation as long as
$\displaystyle{{1\over p}+{1\over q}<{1\over2}}$.

\vtop{
\vspace{-405pt}
\ligne{\hfill\includegraphics[scale=0.7]{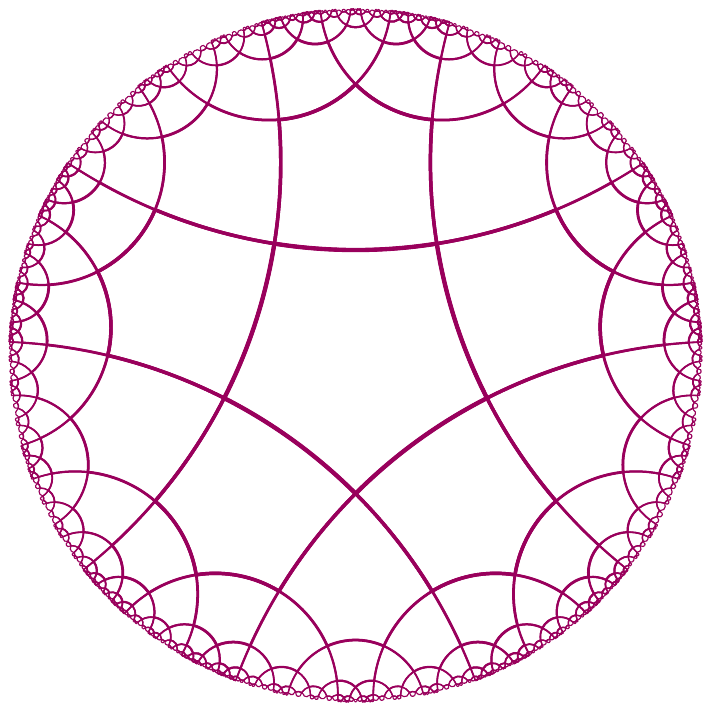}
\hskip -280pt
\includegraphics[scale=0.7]{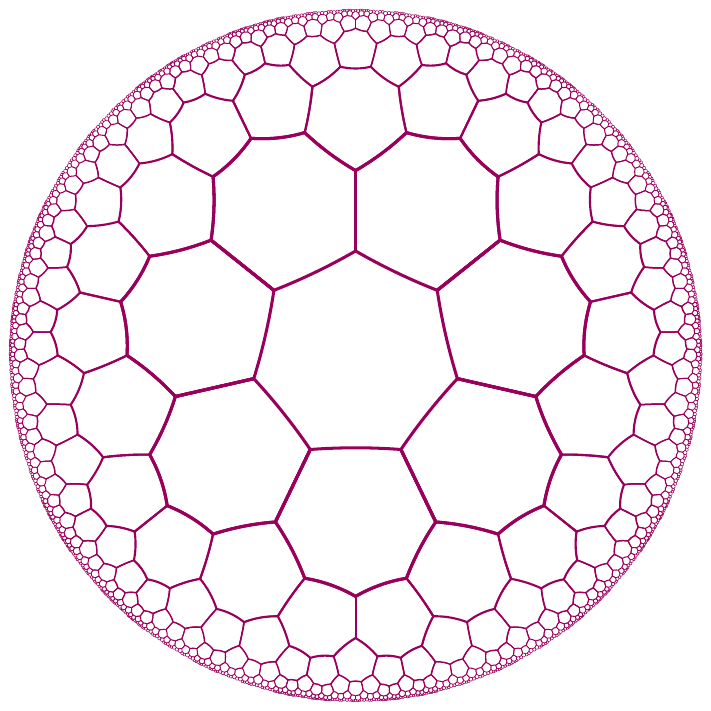}
\hfill}
\vspace{10pt}
\begin{fig}
\label{tilings}
\leurre
Left-hand side: the pentagrid. Right-hand side: the heptagrid.
\end{fig}
}

   We can see that when $q=4$, which corresponds to a right angle for two consecutive edges
of the basis, the smallest number of sides is~5: this is the {\bf pentagrid}, illustrated by
the left-hand side picture of Figure~\ref{poincare}. When $q=3$, which corresponds to
an angle of $\displaystyle{{2\pi}\over3}$ between two consecutive edges of the basis,
the smallest number of sides is~7: this is the {\bf heptagrid}, illustrated by
the right-hand side picture of Figure~\ref{poincare}.

   Note that in the Euclidean plane, the condition for a basis characterized by~$p$ and~$q$
to give rise to a tessellation is 
$\displaystyle{{1\over p}+{1\over q}={1\over2}}$. This gives us finitely many solutions only 
for both~$p$ and~$q$. We have here that $q=3$, 4 or~6 and that then $p=6$, 4 or~3 respectively. 

   From the geometrical point of view, the Chinese project makes use of simple
transformations leaving the tilings glabally invariant: the translations along appropriate
lines, see~\cite{mmbook1,mmbook2,mmbook3}. It also makes use of an appropriate
coordinate system also explained in~\cite{mmbook1,mmbook2,mmbook3}.

\section{Implementing the Chinese keyboard}
\label{implement}

   In this section, Subsection~\ref{basic} introduces the main features of our method and
how it is implemented in various settings: the case of a cell phone, see 
Subsection~\ref{mobile}, and the case of devices equipped with a tactile screen,
see Subsection~\ref{tactile}. We also allow the possiblity to type texts using
a predictive system: thi is described in Subsection~\ref{predictive}. 

\subsection{Basic features}
\label{basic}

    As illustrated by Figure~\ref{pinyin}, we use two disks taken in a tiling, the heptagrid,
which is introduced in Section~\ref{pentahepta}. Indeed, the syllabs of the Chinese languages
have a simple structure. They start with a consonent, possibly empty, then a vowel possibly
ending with~{\bf n}. Note that a few vowels are nasal ones as in the French language.
The left-hand side disc of Figure~\ref{pinyin} gives the possible consonants
of a syllab. The order of the letters follows the traditional orders of consonants for
the pin-yin writing. We can distinguish six groups of letters: 
{\tt b p m f $\vert$ d t n l $\vert$ g k h $\vert$ j q x $\vert$ zh ch sh r $\vert$
z c s}. Two additional consonents, {\tt w} and~{\tt y} are used to write syllabs which
start with the vowels {\tt i} or~{\tt \"u}, introduced by~{\tt y}, and~{\tt u} introduced 
by~{\tt w}. Together with {\tt w} and {\tt y}, in the same sector, we have the three
vowels, {\tt a}, {\tt o} and {\tt e}, indicating the way to access vowels for syllabs
starting with a, o or~e.

\vtop{
\vspace{-220pt}
\ligne{\hfill\includegraphics[scale=0.475]{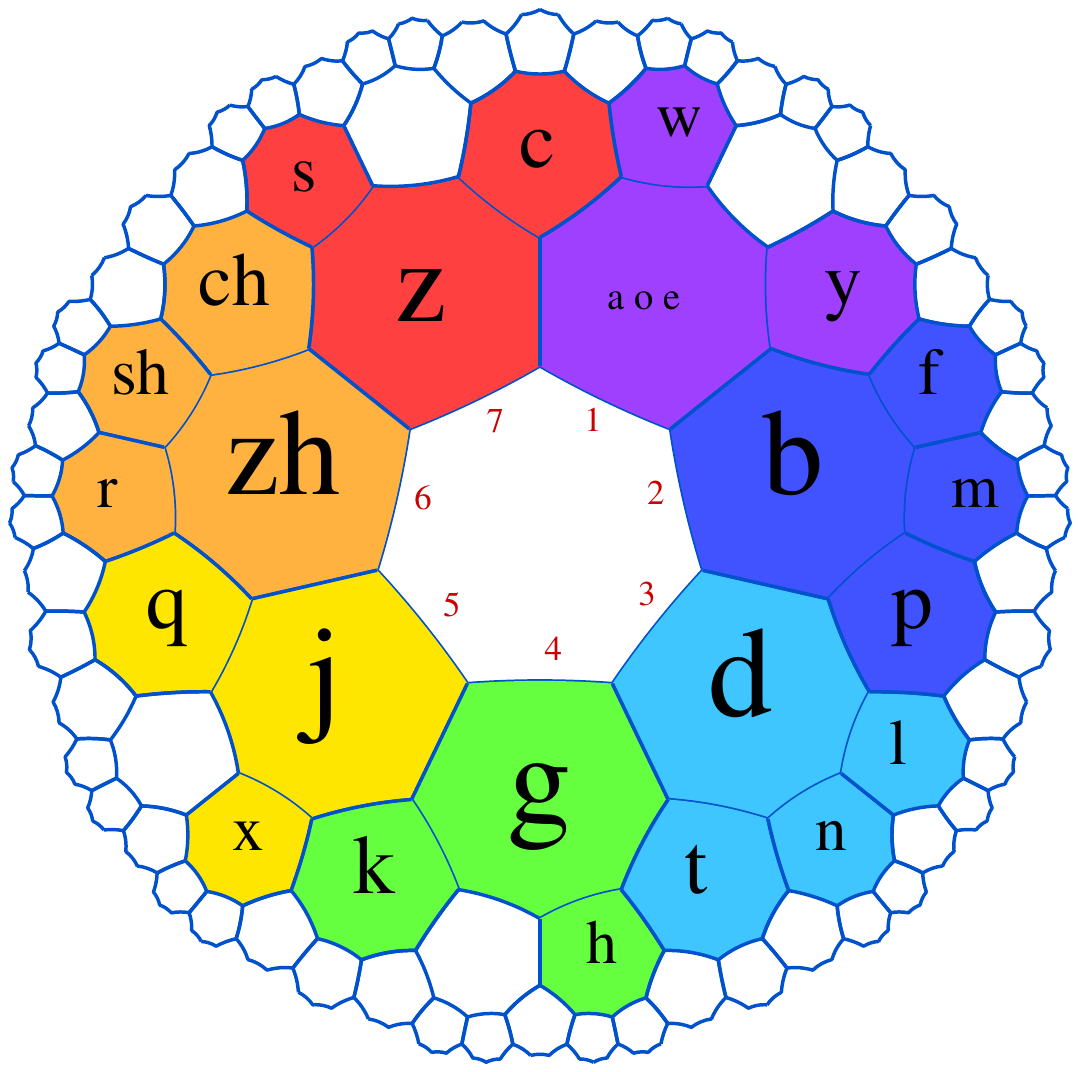}
\hskip -120pt
\includegraphics[scale=0.475]{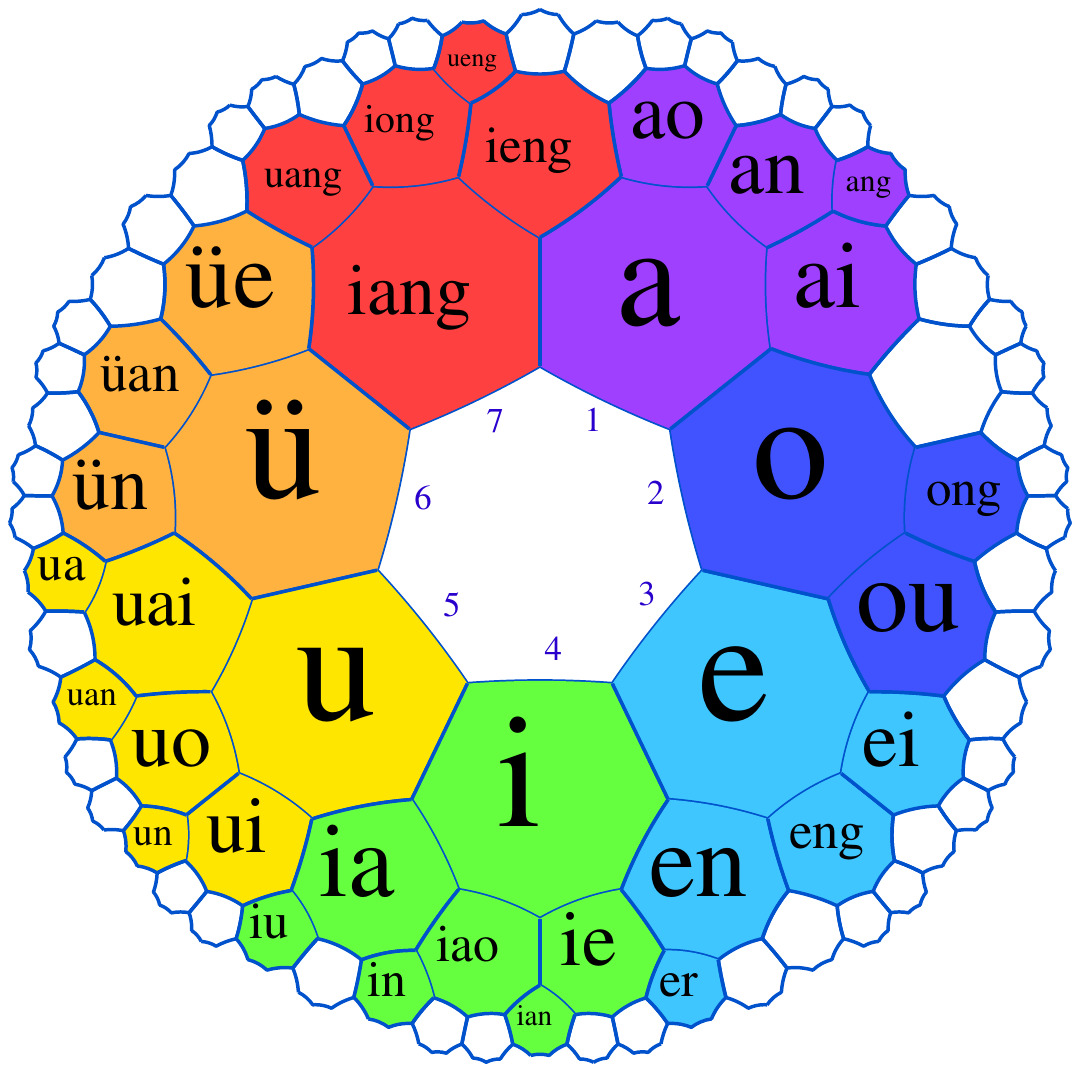}
\hfill}
\vspace{-5pt}
\begin{fig}
\label{pinyin}
\leurre
The display of the consonents and the vowels. Note the cell with {\tt a}, {\tt o} 
and~{\tt e} in the display of the consonents. 
\end{fig}
}

   In a device doted with a tactile screen, the input is direct: the user puts his/her finger
on the appropriate cell to enter the corresponding syllab. 

   In a device with a passive screen, no more than five cliks are needed to enter a syllab. 
Note that with a computer keyboard, the longest syllab requires six letters in pin-yin.
As we have to compare similar devices, note that on a cell phone, a single letters requires 
a bit more than two clicks in mean in order
to be reached. Now, it is not difficult to see that the number of clicks to enter a syllab
in our project is not greater than the number of letters to write it on a computer.

\vtop{
\vspace{-145pt}
\ligne{\hfill\includegraphics[scale=0.325]{bopomofo_7_3.pdf}
\hskip -85pt
\includegraphics[scale=0.325]{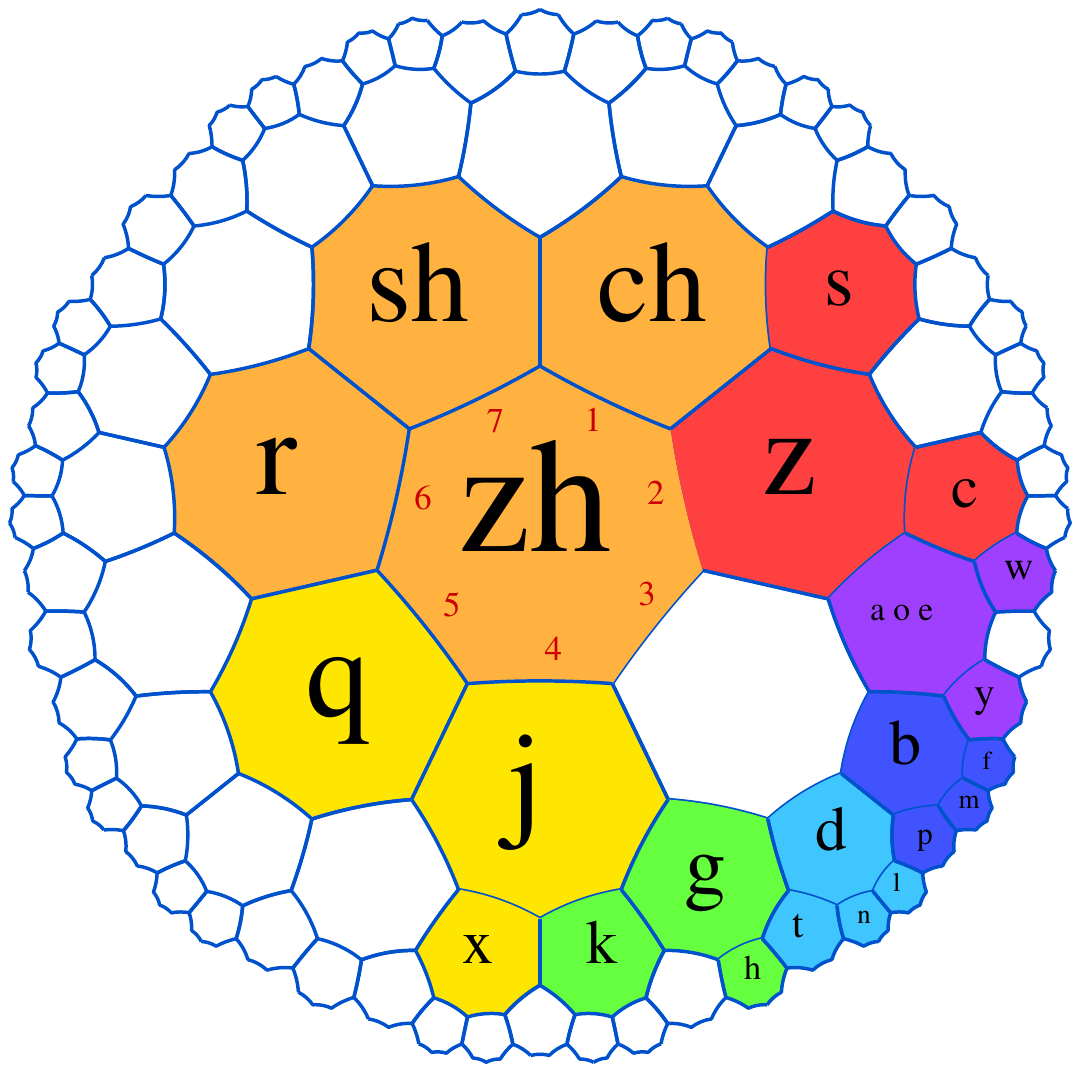}
\hskip -85pt
\includegraphics[scale=0.325]{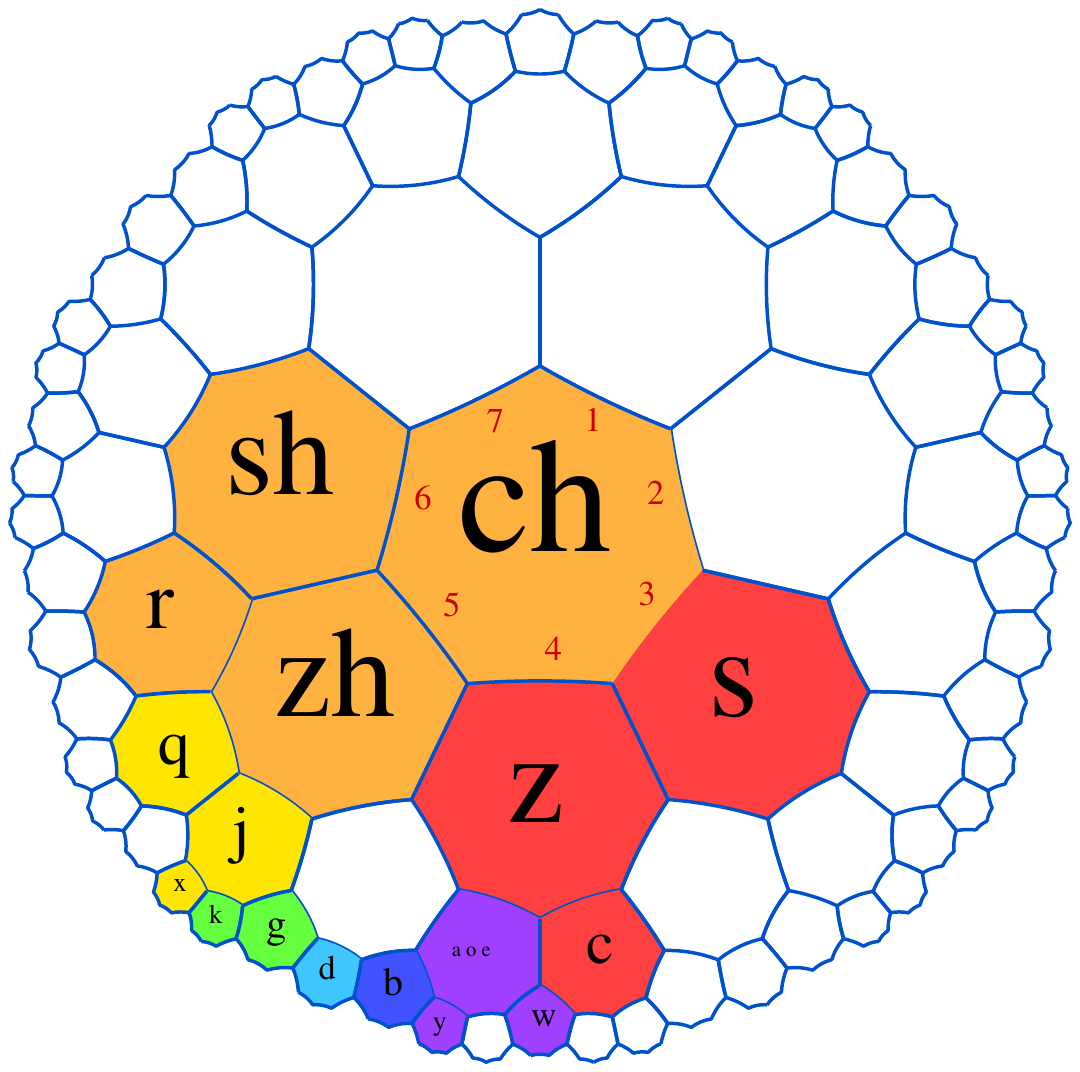}
\hfill}
\vspace{-5pt}
\begin{fig}
\label{conson}
\leurre
Left-hand side: idle configuration. Middle: by pressing key~{\tt 6}, the user
makes {\tt zh} move to the central cell.
Right-hand side: now, by pressing key~{\tt 1}, the user makes {\tt ch} move into the central 
cell, ready to be selected.
\end{fig}
}

   Let us look at the writing of {\tt chuang}, which requires 6 letters. Two cliks allow us to 
put {\tt ch} into the central cell, see Figure~\ref{conson}. Indeed, first, the user
presses on~6 which puts {\tt zh} into the centre. Then, he/she presses on~1 to get
{\tt ch} in the centre. Two cliks also allow us to
put {\tt uang} into the centre, see Figure~\ref{vowel}: first, pressing on~7 puts
{\tt iang} into the centre and then, again pressing on~7 puts {\tt uang} into the centre. 
It is not difficult to manage things in such a way that when the consonent is selected, the 
panel with wowels automatically is presented.

   Note that in the case of a tactile screen, the just mentioned method to input a syllab 
necessitates two gestures only.

\vtop{
\vspace{-145pt}
\ligne{\hfill\includegraphics[scale=0.325]{voyelles_7_3.pdf}
\hskip -85pt
\includegraphics[scale=0.325]{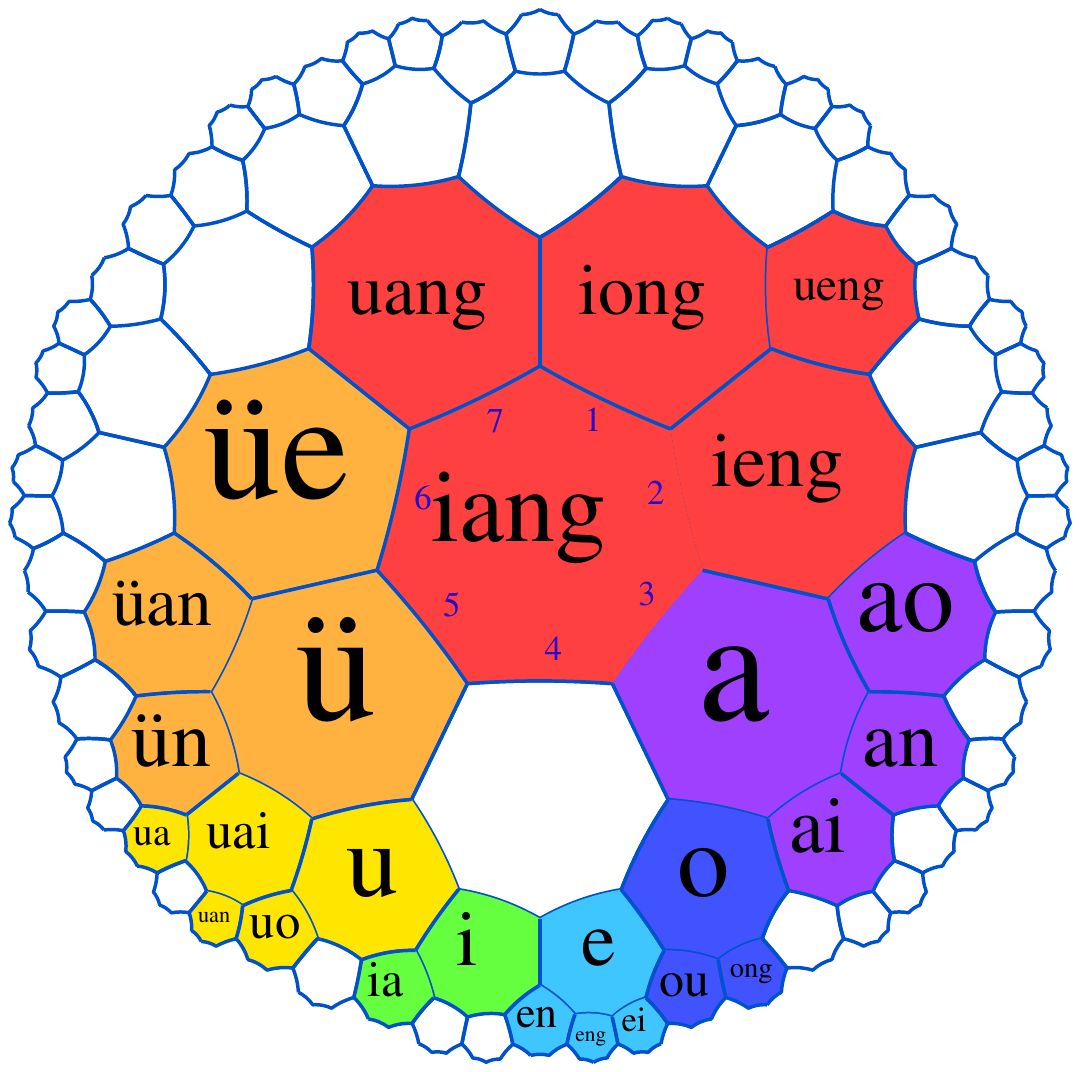}
\hskip -85pt
\includegraphics[scale=0.325]{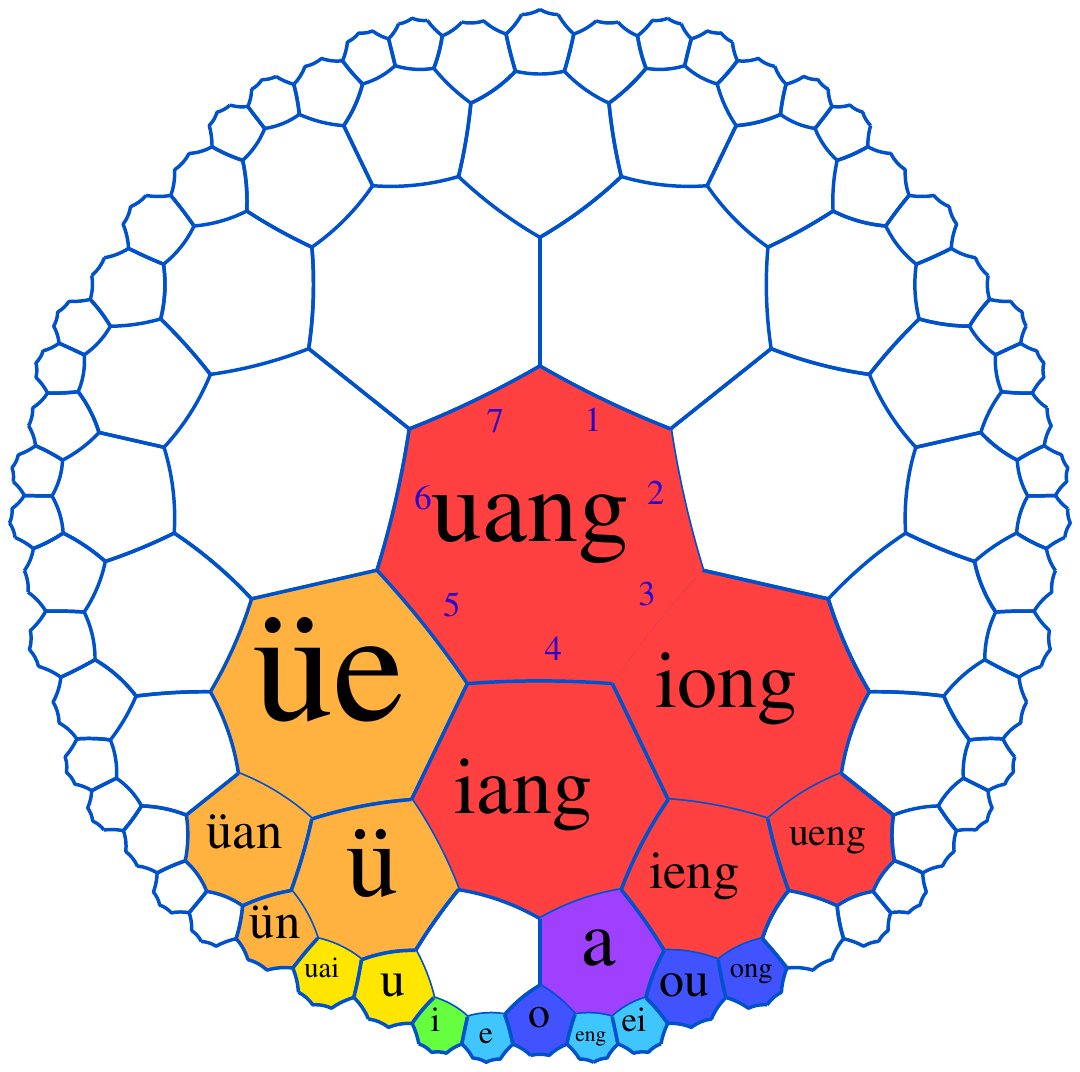}
\hfill}
\vspace{-5pt}
\begin{fig}
\label{vowel}
\leurre
Left-hand side: idle configuration. Middle: {\tt iang} is moved to the central cell.
Right-hand side: {\tt uang} is in the central cell, ready for selection.
\end{fig}
}

   In the same way, we can perform the selection of Chinese characters.

   Once the pin-yin syllab is selected, a panel is displayed with the characters associated 
with this pronouciation which fall within this frame. This time we use the pentagrid for 
readability reasons: the cells in the pentagrid are significantly bigger than in the heptagrid.
The counter-part is that we may display at most 60 characters, while it would be possible to 
display 84 of them in the heptagrid. We shall even display a bit less: we use one sector of
the pentagrid per tone. The fith tone can be used for situations when there are characters
attached to a pronounciation with the neutral tone. But, in most cases, such a display will 
be enough. In some cases, when more than 48 characters must be displayed, a click on an 
approriate key will provide a new plate of characters. Note that compared with the
present tools for typing Chinese texts, this method offers much bigger facility.

\vtop{
\vspace{-20pt}
\ligne{\hfill
\includegraphics[scale=0.3]{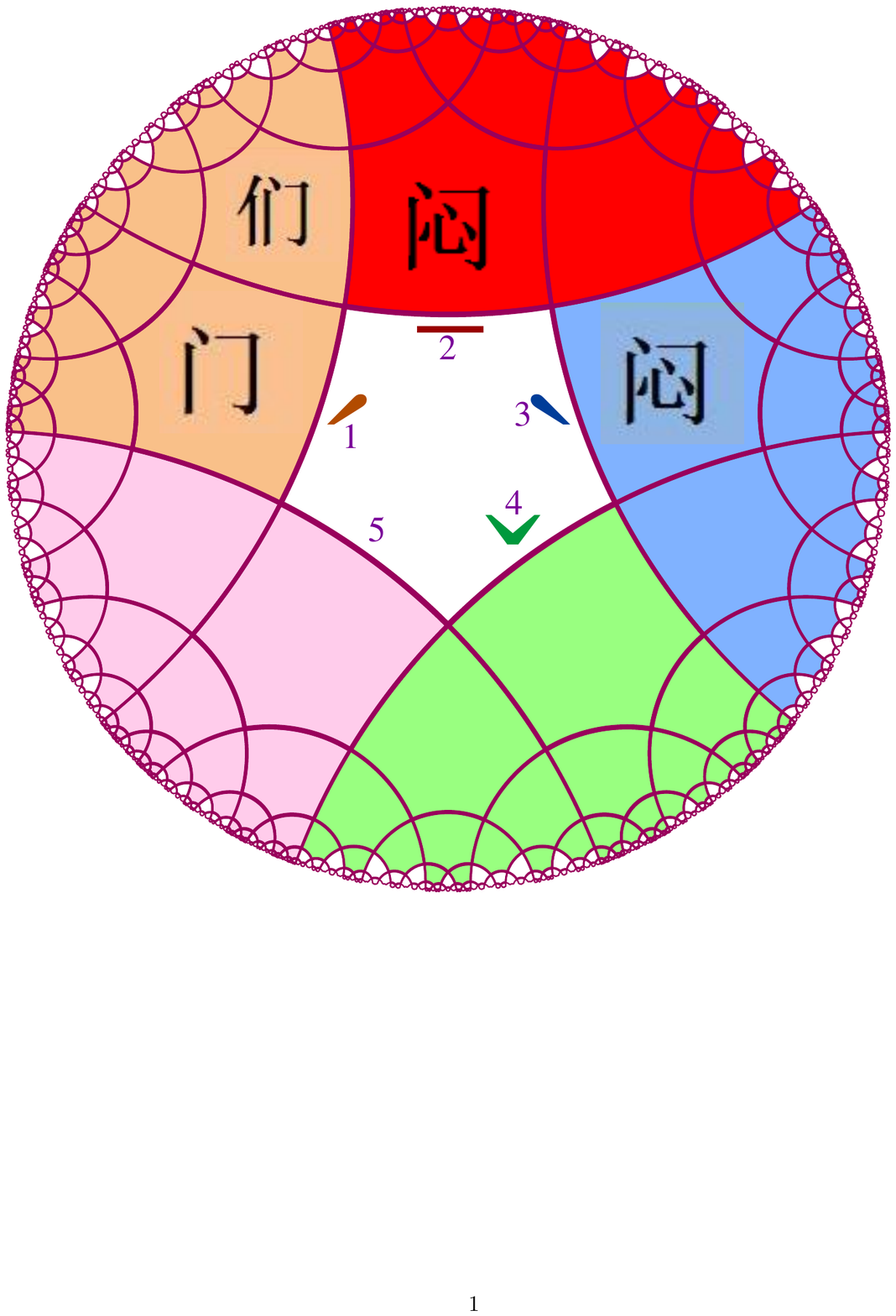}
\hskip -55pt
\includegraphics[scale=0.3]{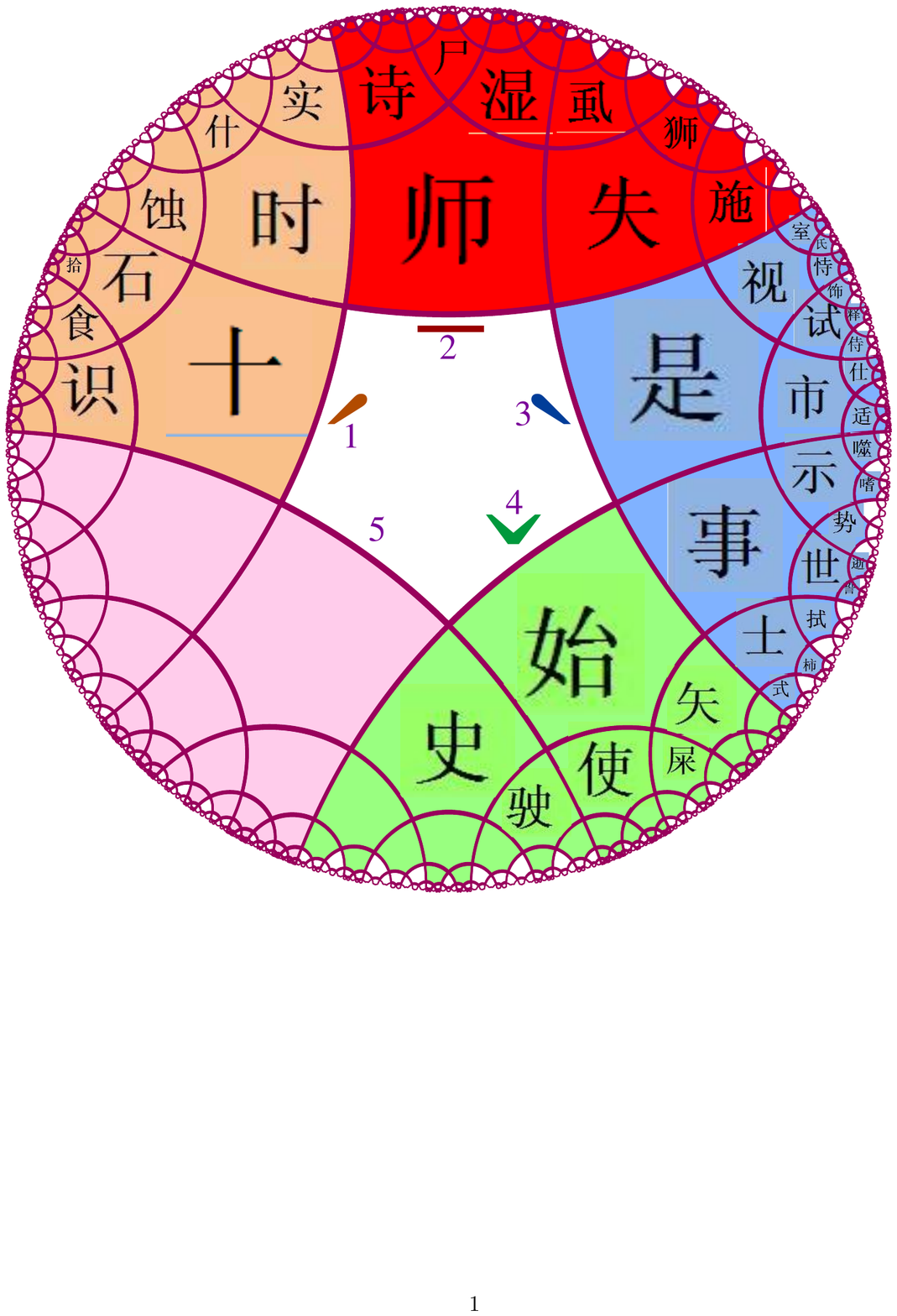}
\hfill}
\vspace{-80pt}
\begin{fig}
\label{testa}
\leurre
Left-hand side: the characters associated with the pin-yin {\tt men}.
Right-hand side: the characters associated with the pin-yin {\tt shi}.
\end{fig}
}

   Figure~\ref{testa} presents two extreme situations for plates of Chinese characters
associated to each pin-yin. On the left-hand side picture, we have the characters
dsiplayed for the pin-yin {\tt men}. On the right-hand side, we have those for the pin-yin
{\tt shi}. In fact, for the pin-yin {\tt men} we need not the two copies of
\hbox{\includegraphics[scale=0.225]{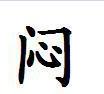}} in our example. But this may be the 
case for other 
characters which may be associated to the same pin-yin but different tones. This is not the case
for most characters, but the number of those for which it is the case is not so few that 
we may stumble on them from time to time.

   Now, in order to look better at how the software should work, let us look at the following
example. Suppose that we would like to write

\ligne{\hfill
$^{\grave{}}$\includegraphics[scale=0.25]{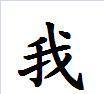}
\hskip-7pt
\includegraphics[scale=0.25]{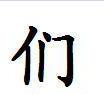}
\hskip-7pt
\includegraphics[scale=0.25]{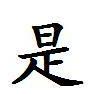}
\hskip-7pt
\includegraphics[scale=0.25]{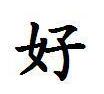}
\hskip-7pt
\includegraphics[scale=0.25]{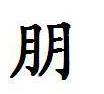}
\hskip-7pt
\includegraphics[scale=0.25]{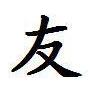}$^{\acute{}}$
\hfill}

    We shall separately consider the case of a cell phone and that of a device with a 
tactile screen.

\subsection{The example on a cell phone}
\label{mobile}

    In Sub-subsection~\ref{protocol}, we present the general protocol which seems to us the 
most accessible to a user. In Sub-subsection~\ref{improve}, we introduce two options which
can be of help for the user.

\subsubsection{The general protocol}
\label{protocol}

    Let us illustrate how our project works to enter 
\hbox{\hskip-3pt\includegraphics[scale=0.15]{wo0_moi.jpg}}. 
First, the consonent panel appears as in Figure~\ref{panwo}. In  the figure, we
have a reproduction of the outline of the cellphone and how the heptagrid appears in it.
The keyboard is displayed on the right-hand side of the window devoted to the dipslay of the
grids. There are 14 keys on the keyboard. Seven of them allow the user to put the required
cell on the center of the pentagrid. The other keys may have a use, as later indicated.

   Then the figure moves as shown by Figure~\ref{panwo1} in order to get the display shown
by Figure~\ref{panwo2}.  The user presses on~{\tt 1} as shown by Figure~\ref{panwo1}.
The user presses {\tt 1} again and then {\tt ok}. This makes the display of Figure~\ref{panwo2}
appear on the screen. This triggers the display of the vowel panel, see
Figure~\ref{vowwo1}. Here, the vowel panel gives an immediate access to the appropriate
vowel, namely~{\tt o}, see Figure~\ref{vowwo2}. This makes the panel of characters whose pin-yin
is {\tt wo} appear, see Figure~\ref{thewo1}. There are more characters than those indicated
in Figure~\ref{thewo1} but, as done in most softwares, the panel of the figure shows the most
used of them. 

Now, the access of the expected character is also immediate as it is in a tile which is
immediately around the central tile. Note that the panel of characters is now displayed in a copy
of the pentagrid: the tiles are bigger corresponding to the size of the Chinese characters, 
usually bigger than roman letters.

   Also, note that the pentagrid make use of four sectors only: this due to the fact that
each sector contains the characters which are pronounced in the same tone. 

\vtop{
\vspace{-150pt}
\hbox{\hskip 50pt\includegraphics[scale=0.35]{bopomofo_7_3.pdf}}
\vspace{-390pt}
\hbox{\hskip 30pt\includegraphics[scale=0.6]{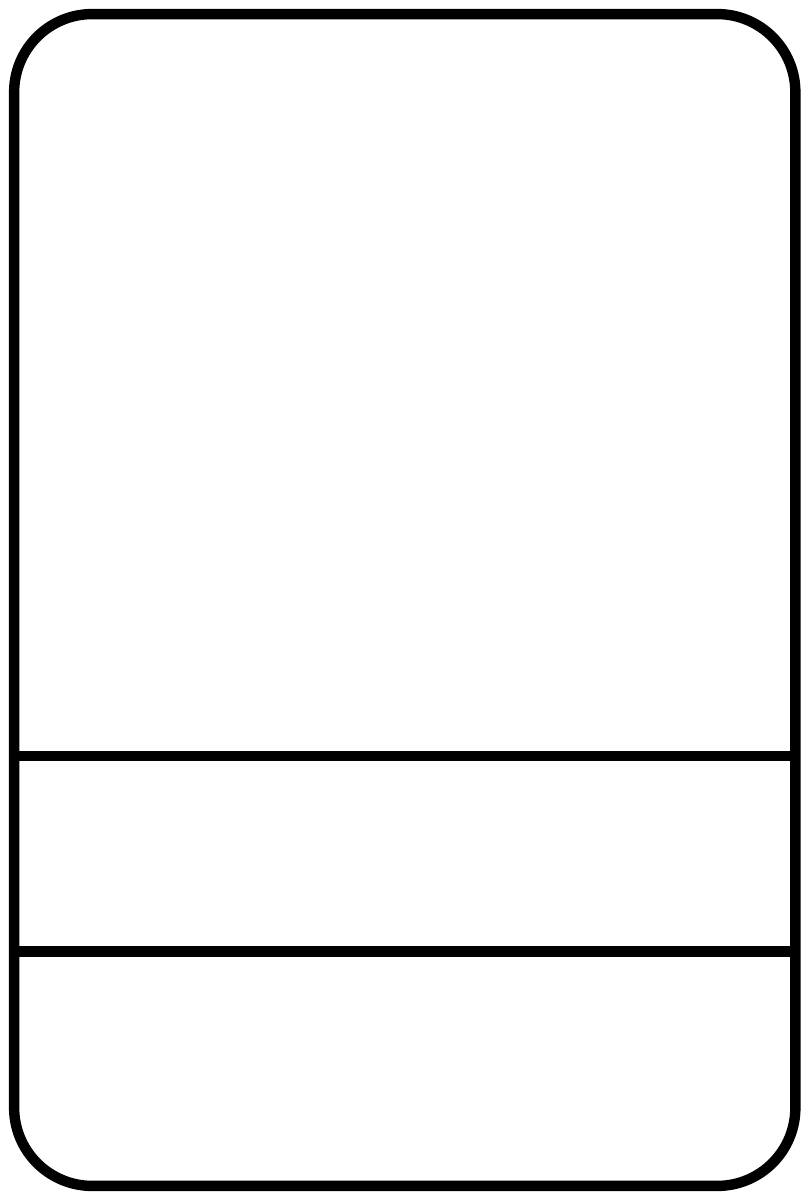}}
\vspace{-370pt}
\hbox{\hskip 215pt\includegraphics[scale=0.4]{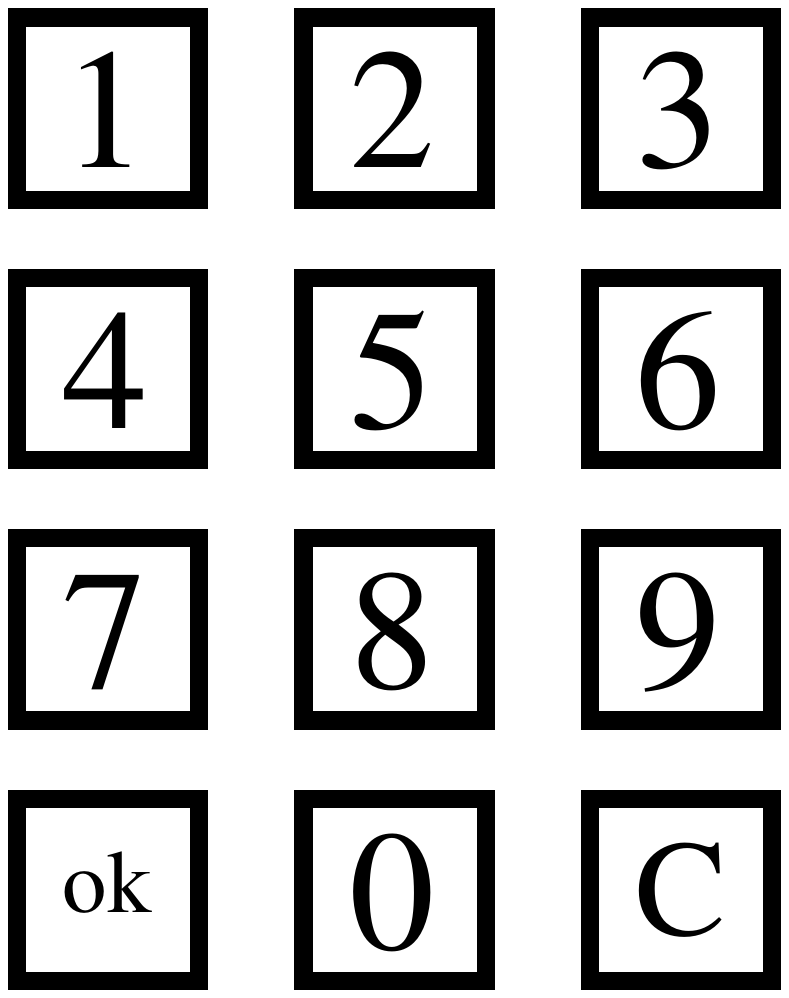}}
\vspace{45pt}
\begin{fig}\label{panwo}
First, the consonent panel.
\end{fig}
}

\vskip 15pt

\vtop{
\vspace{-150pt}
\hbox{\hskip 50pt\includegraphics[scale=0.35]{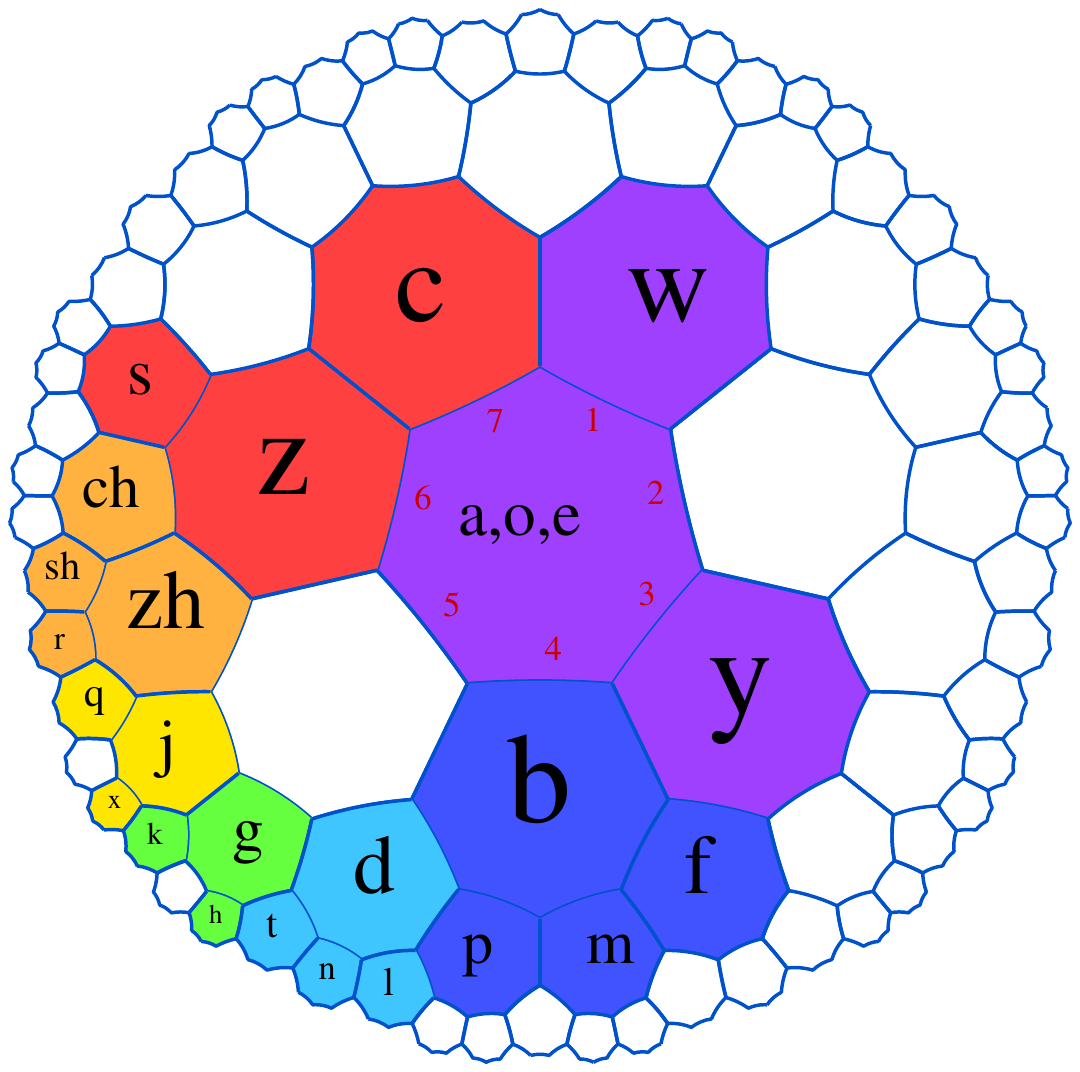}}
\vspace{-390pt}
\hbox{\hskip 30pt\includegraphics[scale=0.6]{cadre.pdf}}
\vspace{-370pt}
\hbox{\hskip 215pt\includegraphics[scale=0.4]{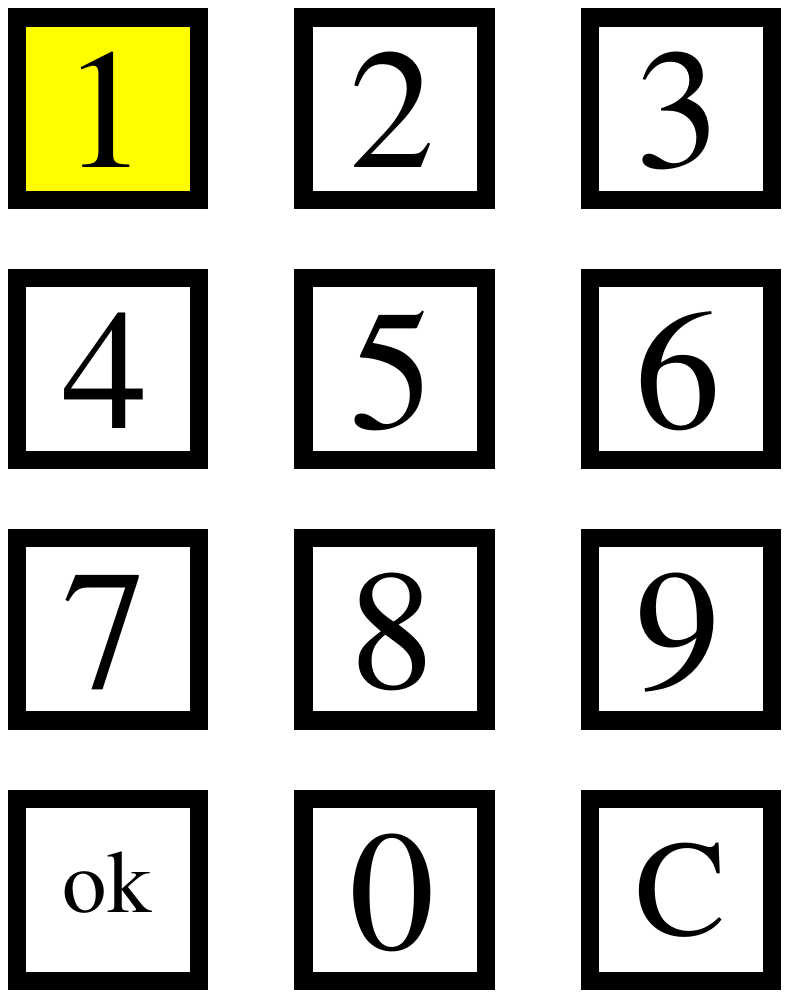}}
\vspace{45pt}
\begin{fig}\label{panwo1}
Choosing the letter.
\end{fig}
}

\vtop{
\vspace{-150pt}
\hbox{\hskip 50pt\includegraphics[scale=0.35]{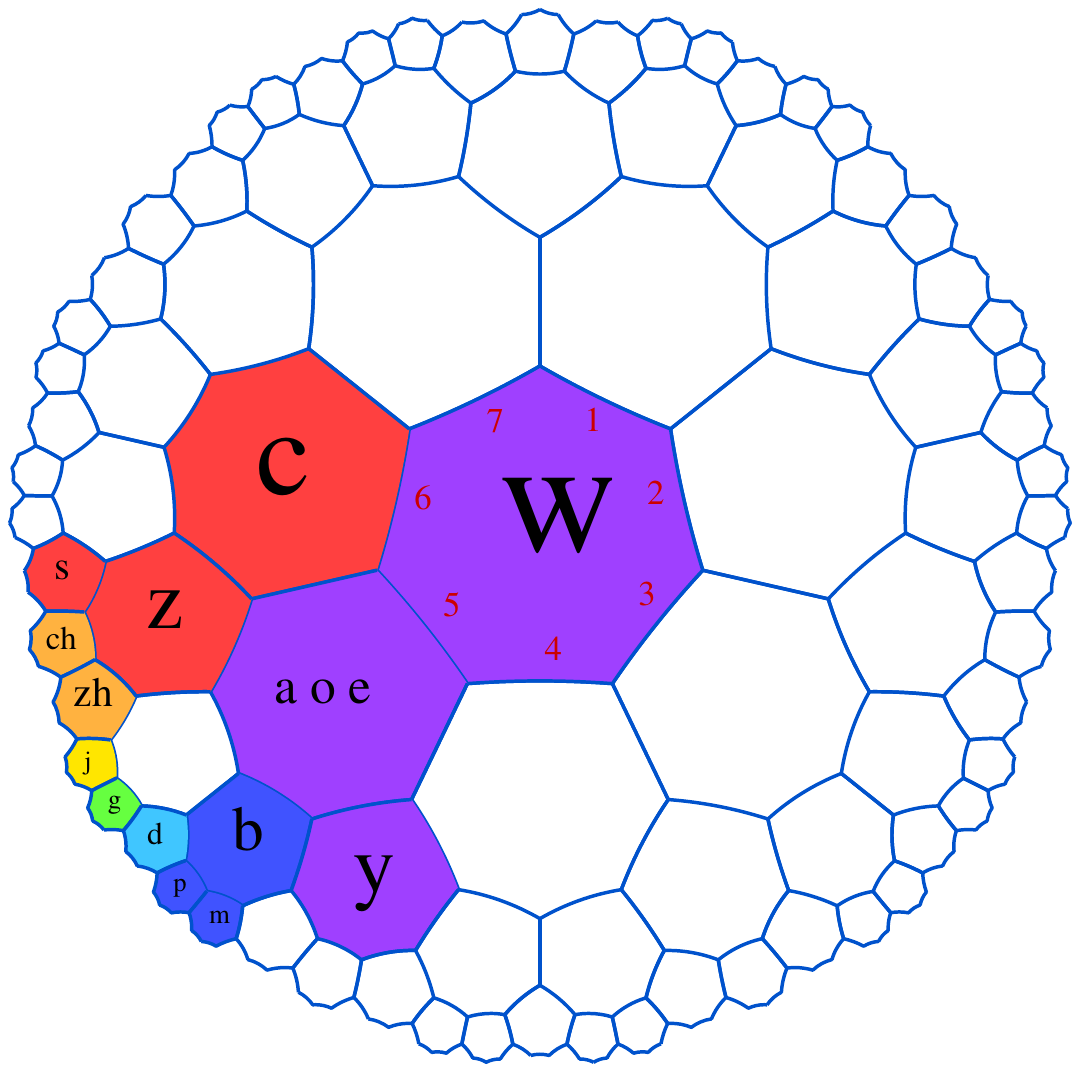}}
\vspace{-390pt}
\hbox{\hskip 30pt\includegraphics[scale=0.6]{cadre.pdf}}
\vspace{-370pt}
\hbox{\hskip 215pt\includegraphics[scale=0.4]{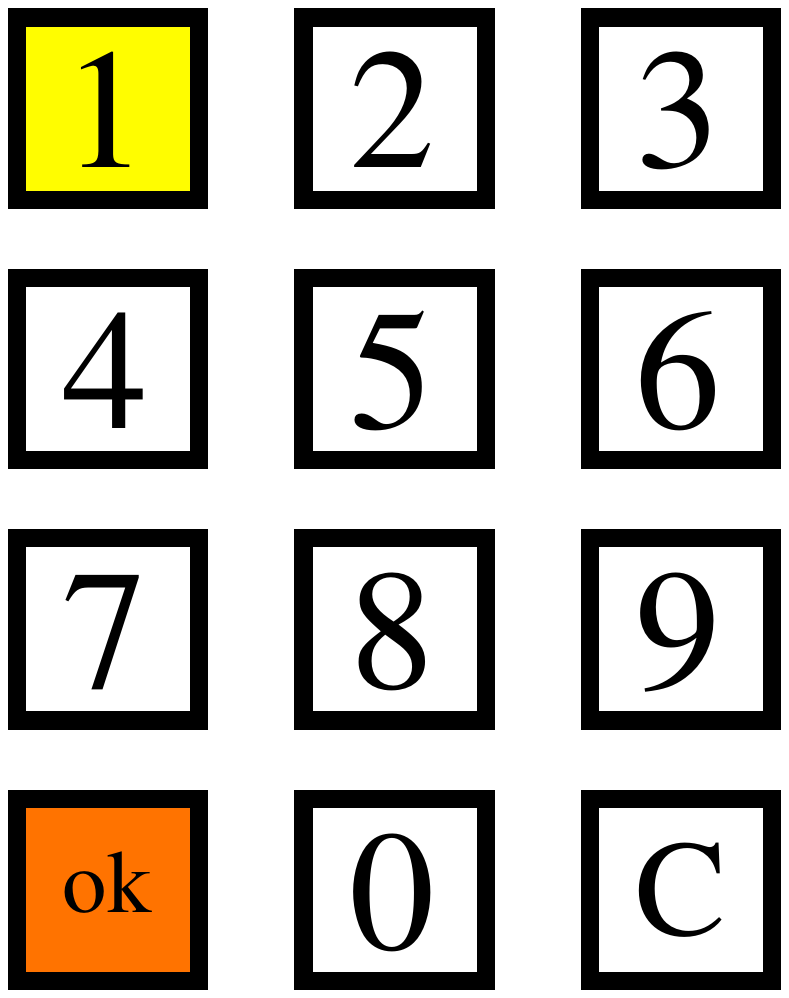}}
\vspace{-32.5pt}
\hbox{\hskip 50pt\tt w}
\vspace{67.5pt}
\begin{fig}\label{panwo2}
The first letter is choiced, here {\tt w}.
\end{fig}
}

\vskip 20pt
\vtop{
\vspace{-150pt}
\hbox{\hskip 50pt\includegraphics[scale=0.35]{voyelles_7_3.pdf}}
\vspace{-390pt}
\hbox{\hskip 30pt\includegraphics[scale=0.6]{cadre.pdf}}
\vspace{-370pt}
\hbox{\hskip 250pt\includegraphics[scale=0.4]{clavier.pdf}}
\vspace{-32.5pt}
\hbox{\hskip 50pt\tt w}
\vspace{67.5pt}
\begin{fig}\label{vowwo1}
The panel of vowels is triggered.
\end{fig}
}

\vtop{
\vspace{-150pt}
\hbox{\hskip 50pt\includegraphics[scale=0.35]{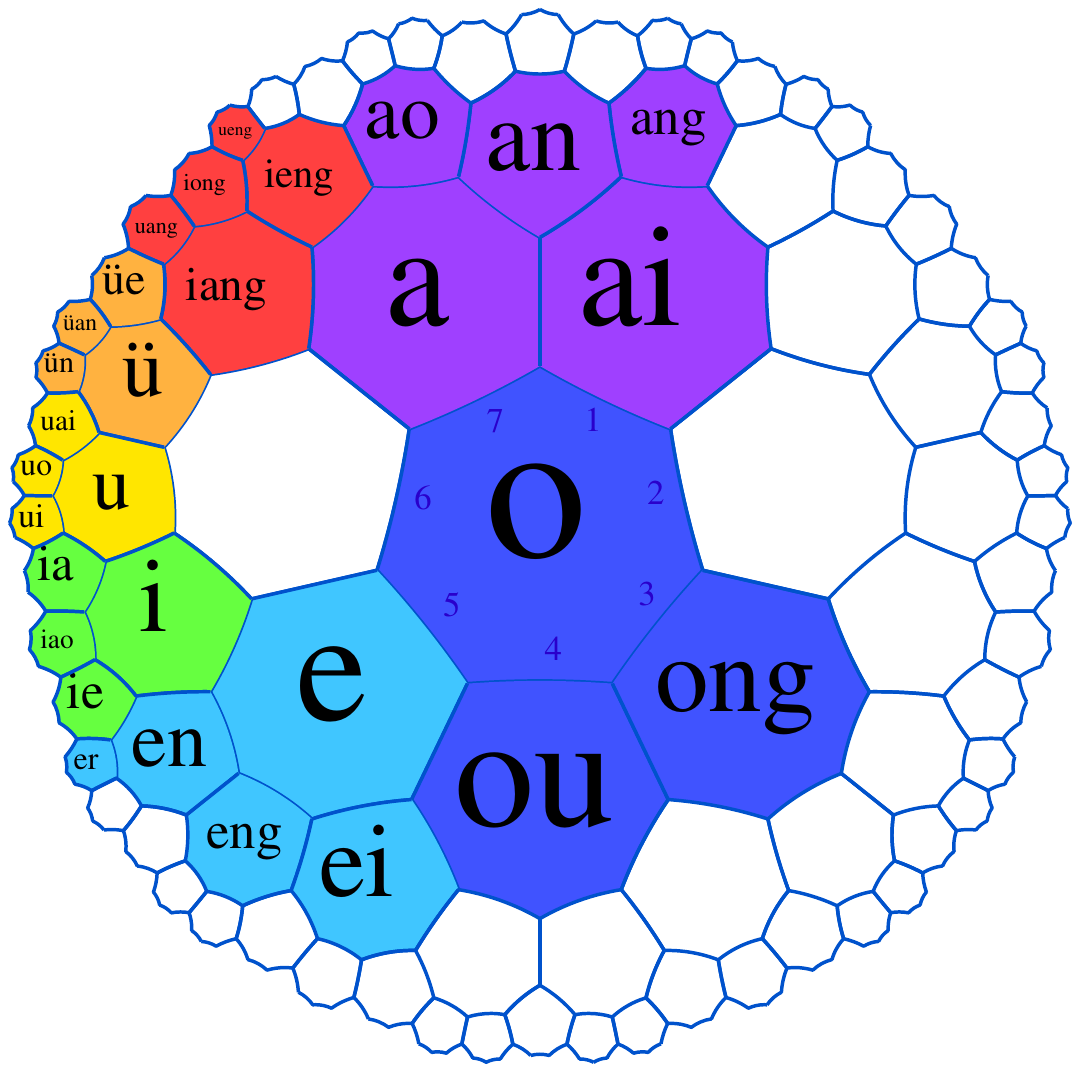}}
\vspace{-390pt}
\hbox{\hskip 30pt\includegraphics[scale=0.6]{cadre.pdf}}
\vspace{-370pt}
\hbox{\hskip 250pt\includegraphics[scale=0.4]{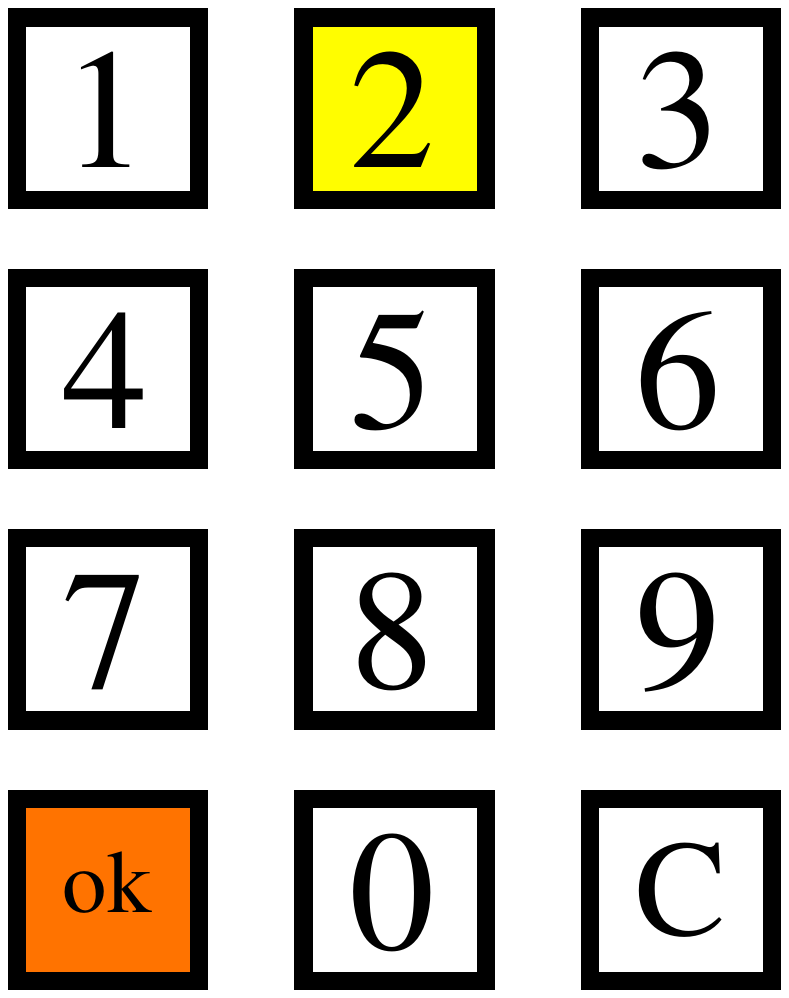}}
\vspace{-32.5pt}
\hbox{\hskip 50pt\tt wo}
\vspace{67.5pt}
\begin{fig}\label{vowwo2}
The expected vowel is selected, here {\tt o}, so that {\tt wo} is selected.
\end{fig}
}

\vtop{
\hbox{\hskip 30pt\includegraphics[scale=0.275]{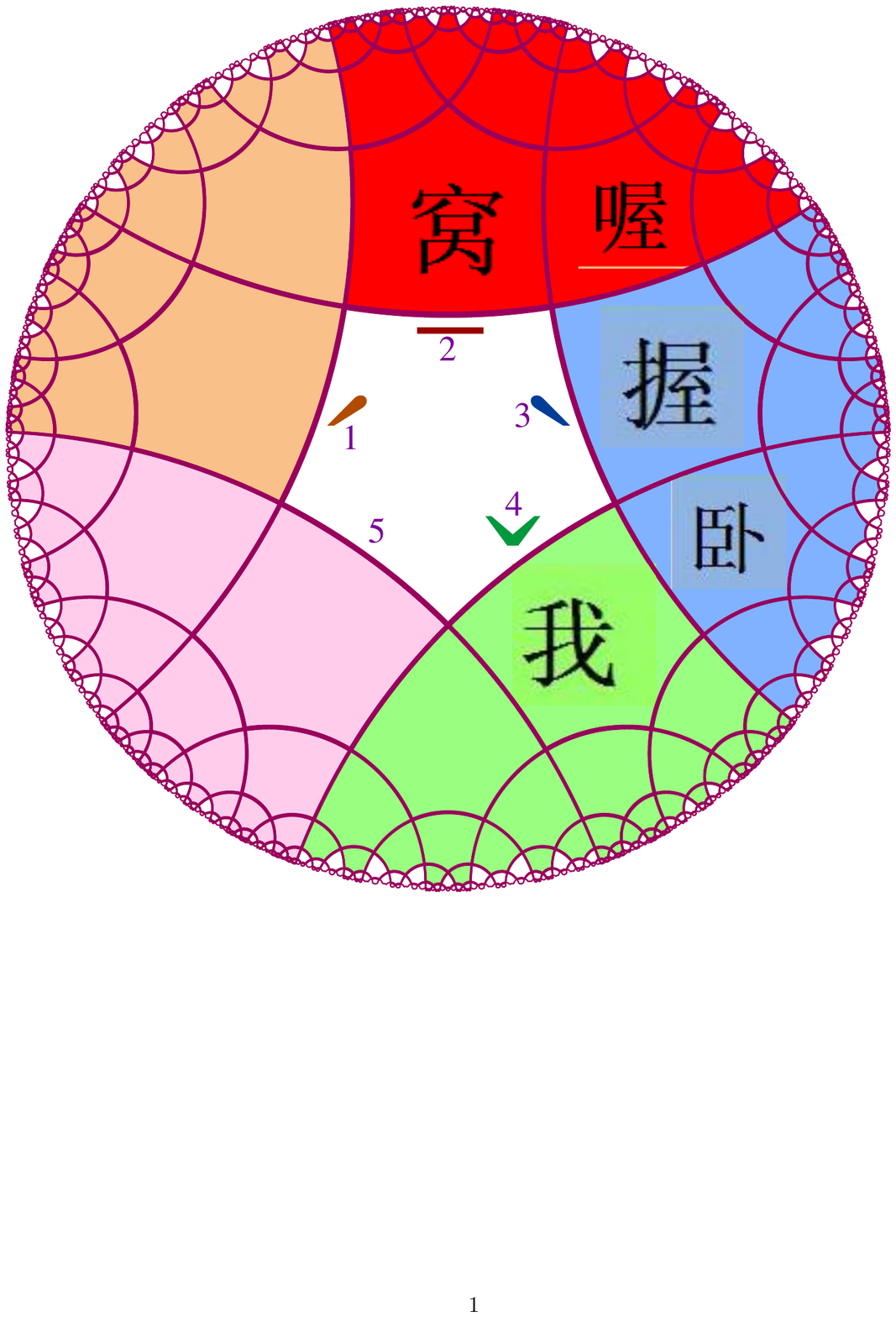}}
\vspace{-465pt}
\hbox{\hskip 30pt\includegraphics[scale=0.6]{cadre.pdf}}
\vspace{-370pt}
\hbox{\hskip 250pt\includegraphics[scale=0.4]{clavier.pdf}}
\vspace{-32.5pt}
\hbox{\hskip 50pt\tt wo}
\vspace{67.5pt}
\begin{fig}\label{thewo1}
The panel of the characters whose pin-yin is {\tt wo} appears.
\end{fig}
}
\vskip 20pt

This display
should help the user who, in principle, knows under which tone the expected character is
pronounced. In order to facilitate the location of the zone to which the character is to be 
found, the same colours are given to each sector. Accordingly, we decided that
the first,second, third and fourth tone is red, orange, green and blue respectively,
see Figures~\ref{thewo1}, \ref{thewo2}, \ref{theshi1} and~\ref{theshi2}. Once the 
character is selected, it appears on the {\bf control screen}, which appears below
the screen of the panels, see the just mentioned figures.

   We give again another example for the same sentence with the syllab {\tt shi}. The selection
of this syllab involves several points which have to be seen too.

\vskip 15pt
\vtop{
\vspace{-20pt}
\hbox{\hskip 30pt\includegraphics[scale=0.275]{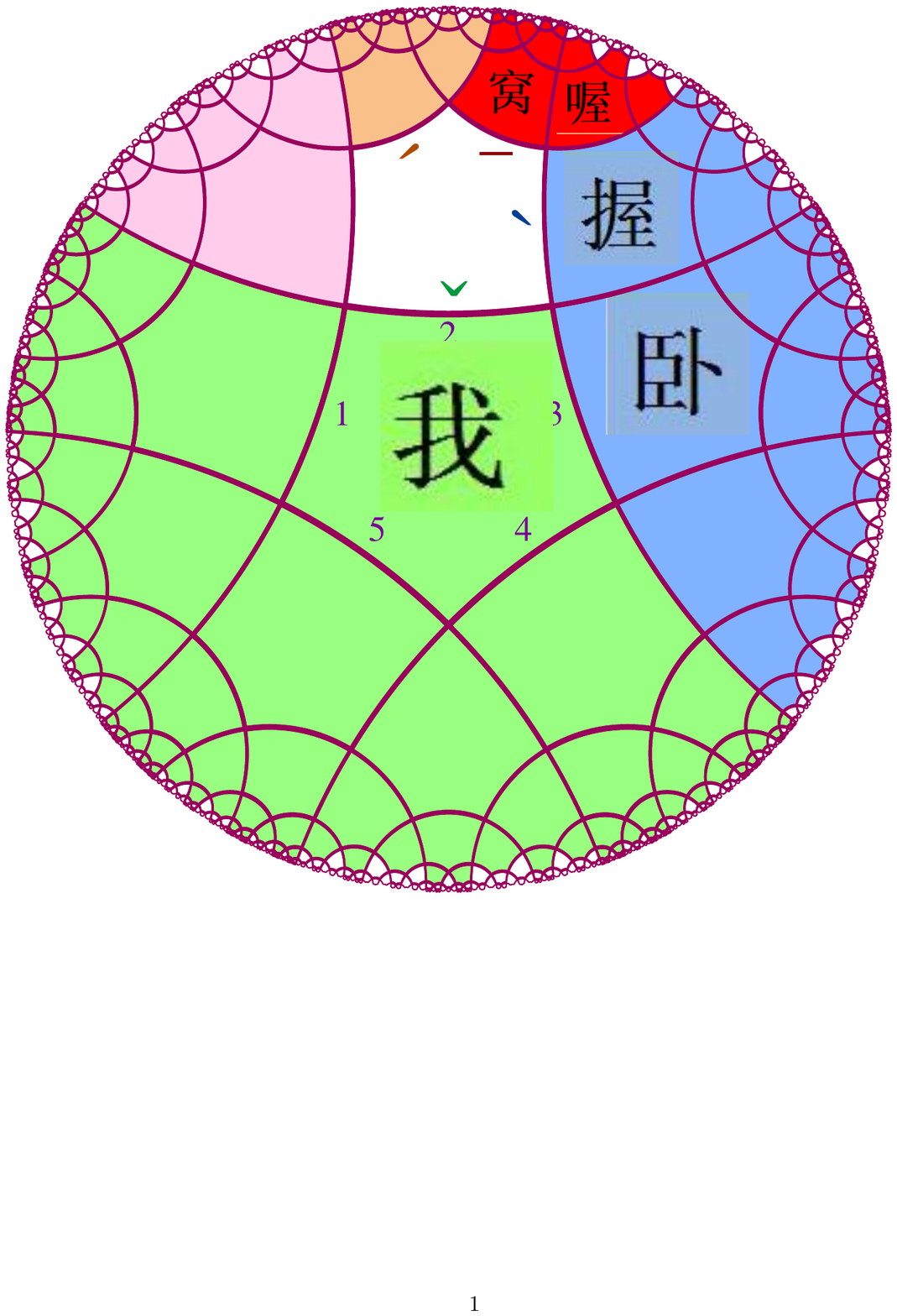}}
\vspace{-465pt}
\hbox{\hskip 30pt\includegraphics[scale=0.6]{cadre.pdf}}
\vspace{-370pt}
\hbox{\hskip 250pt\includegraphics[scale=0.4]{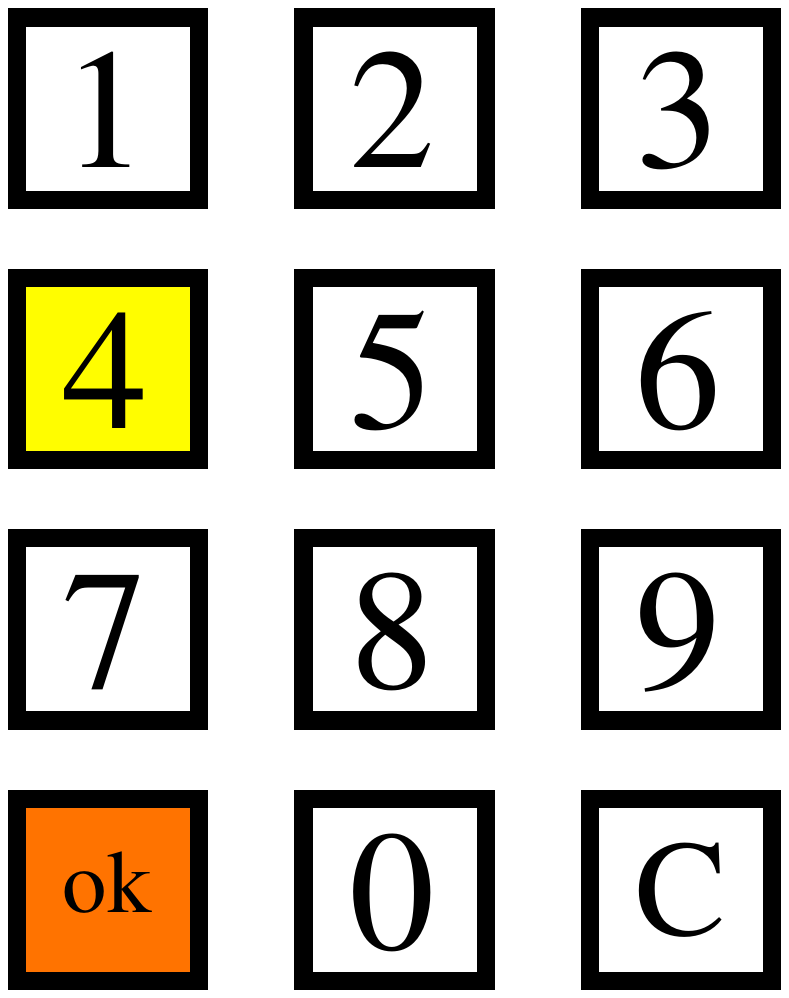}}
\vspace{-27.5pt}
\hbox{\hskip 15pt\hbox{\hskip 40pt\includegraphics[scale=0.25]{wo0_moi.jpg}}
}
\vspace{55pt}
\begin{fig}\label{thewo2}
The right character is choosed, here \includegraphics[scale=0.20]{wo0_moi.jpg}.
\end{fig}
}
\vskip 20pt
\vtop{
\vspace{-150pt}
\hbox{\hskip 50pt\includegraphics[scale=0.35]{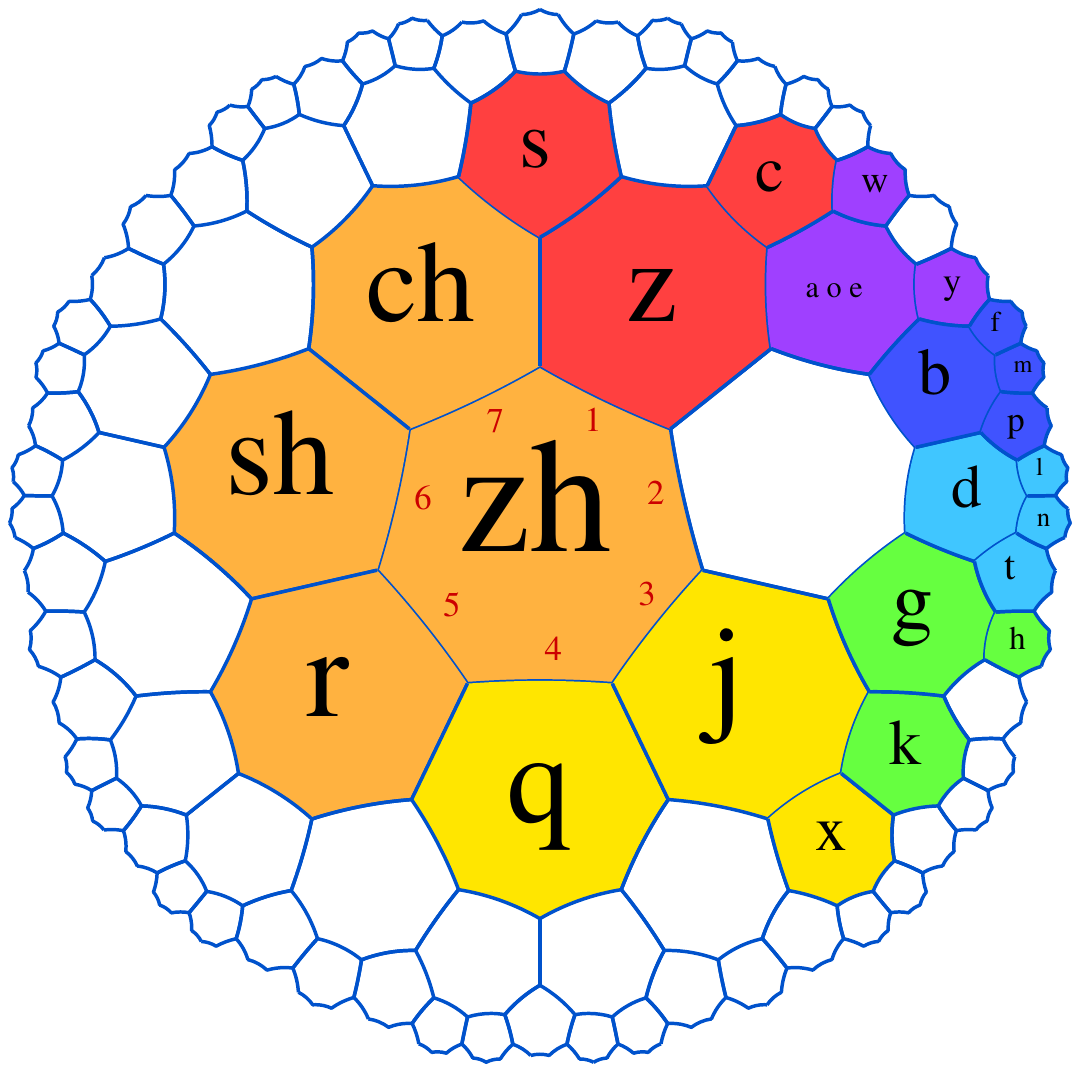}}
\vspace{-390pt}
\hbox{\hskip 30pt\includegraphics[scale=0.6]{cadre.pdf}}
\vspace{-370pt}
\hbox{\hskip 215pt\includegraphics[scale=0.4]{clavier1.pdf}}
\vspace{-27.5pt}
\hbox{\hskip 15pt\hbox{\hskip 40pt\includegraphics[scale=0.25]{wo0_moi.jpg}}
\hskip-27.5pt\hbox{\hskip 20pt\includegraphics[scale=0.25]{men0_pluriel.jpg}}
}
\vspace{55pt}
\begin{fig}\label{panshi1}
Choosing the letter.
\end{fig}
}

   The selection of~{\tt shi} raises a first problem with the consonent {\tt sh}.
Phonetically, it is a consonent as {\tt w} or {\tt s} are. However, in pin-yin it is
written {\tt sh}, so that the user may have in mind that two letters must be issued and not
a single one. We indicate in Sub-subsection~\ref{improve} how to help the user in this
case.

\vtop{
\vspace{-150pt}
\hbox{\hskip 50pt\includegraphics[scale=0.35]{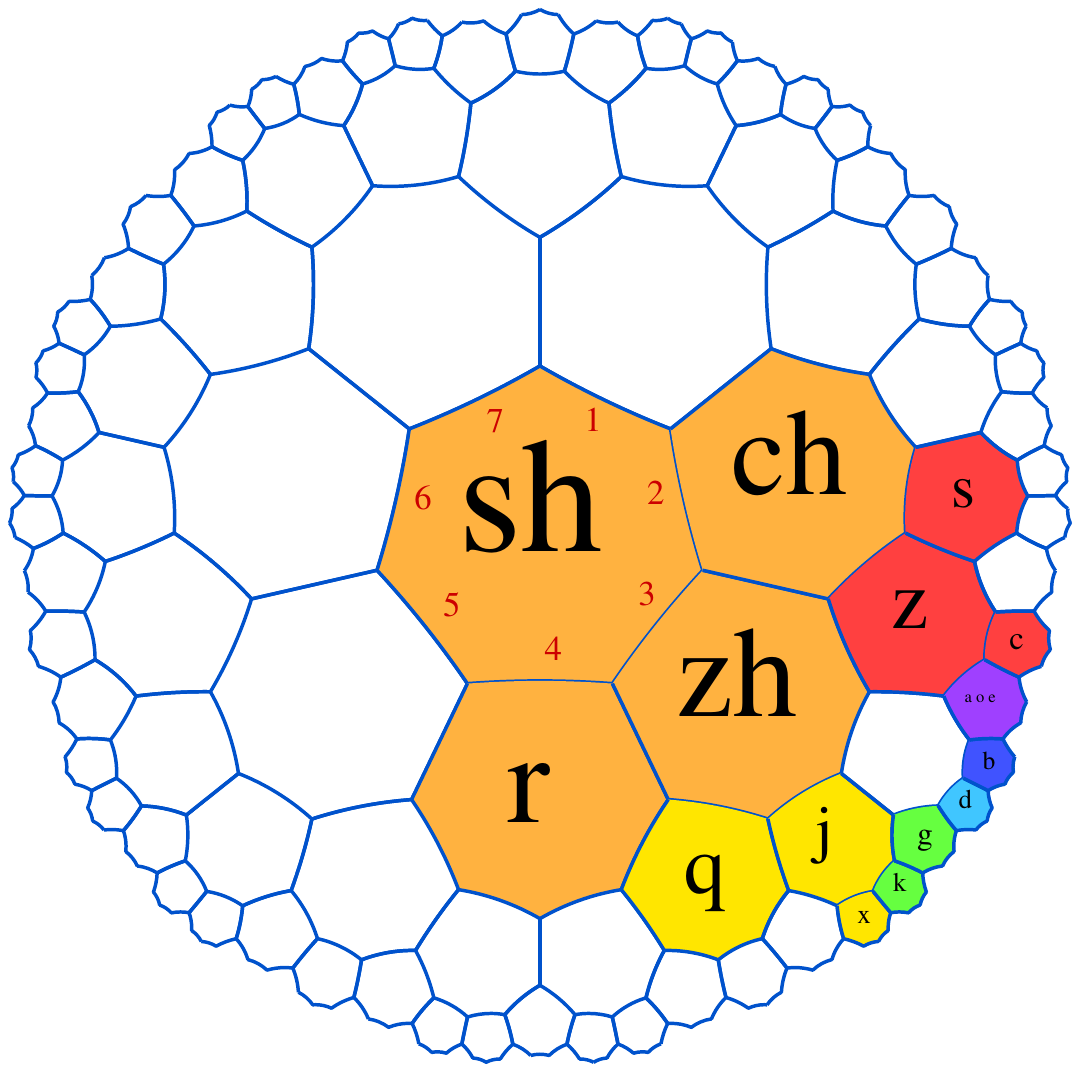}}
\vspace{-390pt}
\hbox{\hskip 30pt\includegraphics[scale=0.6]{cadre.pdf}}
\vspace{-370pt}
\hbox{\hskip 215pt\includegraphics[scale=0.4]{clavier1ok.pdf}}
\vspace{-27.5pt}
\hbox{\hskip 15pt\hbox{\hskip 40pt\includegraphics[scale=0.25]{wo0_moi.jpg}}
\hskip-27.5pt\hbox{\hskip 20pt\includegraphics[scale=0.25]{men0_pluriel.jpg}}
}
\vspace{-27pt}
\hbox{\hskip 50pt\tt sh}
\vspace{73pt}
\begin{fig}\label{panshi2}
The first letter is choiced, here {\tt sh}.
\end{fig}
}

\vtop{
\vspace{-150pt}
\hbox{\hskip 50pt\includegraphics[scale=0.35]{voyelles_7_3.pdf}}
\vspace{-390pt}
\hbox{\hskip 30pt\includegraphics[scale=0.6]{cadre.pdf}}
\vspace{-370pt}
\hbox{\hskip 250pt\includegraphics[scale=0.4]{clavier.pdf}}
\vspace{-27.5pt}
\hbox{\hskip 15pt\hbox{\hskip 40pt\includegraphics[scale=0.25]{wo0_moi.jpg}}
\hskip-27.5pt\hbox{\hskip 20pt\includegraphics[scale=0.25]{men0_pluriel.jpg}}
}
\vspace{-27pt}
\hbox{\hskip 50pt\tt sh}
\vspace{73pt}
\begin{fig}\label{vowshi1}
The panel of vowels appears.
\end{fig}
}

   The second point is the fact that a lot of characters have the same pin-yin {\tt shi}.
Moreover, the distribution between the tones is not uniform. An overwhelming majority of
characters pronouned~{\tt shi} are said in the fourth tone. This can be seen in the panel
associated to~{\tt shi}. Note that the number of characters presented here is rather small
compared to the whole number of characters associated to~{\tt shi}. Here we offer 40~of them
while their number is over~300.

\vtop{
\vspace{-150pt}
\hbox{\hskip 50pt\includegraphics[scale=0.35]{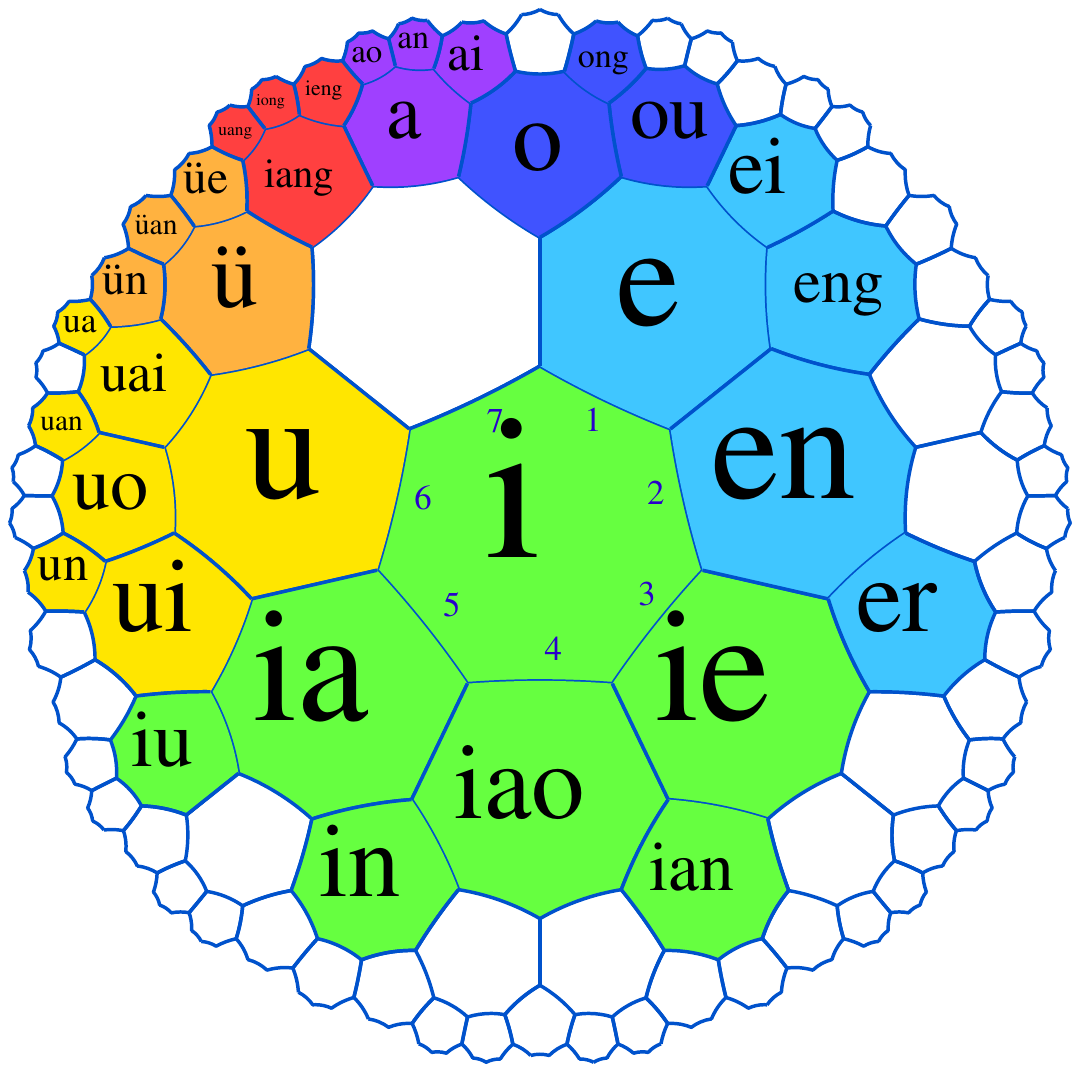}}
\vspace{-390pt}
\hbox{\hskip 30pt\includegraphics[scale=0.6]{cadre.pdf}}
\vspace{-370pt}
\hbox{\hskip 250pt\includegraphics[scale=0.4]{clavier4ok.pdf}}
\vspace{-27.5pt}
\hbox{\hskip 15pt\hbox{\hskip 40pt\includegraphics[scale=0.25]{wo0_moi.jpg}}
\hskip-27.5pt\hbox{\hskip 20pt\includegraphics[scale=0.25]{men0_pluriel.jpg}}
}
\vspace{-27pt}
\hbox{\hskip 50pt\tt shi}
\vspace{73pt}
\begin{fig}\label{vowshi2}
The appropriate vowel is chosen, here {\tt i}.
\end{fig}
}

The choice of the character for~{\tt you} requires an additional
screen as the expected character is not an immediate neighbour of the central tile. This
is also the case for~{\tt men}, although the number of characters for this pin-yin is
very low.

\vtop{
\vspace{-20pt}
\hbox{\hskip 30pt\includegraphics[scale=0.275]{test_shi.pdf}}
\vspace{-465pt}
\hbox{\hskip 30pt\includegraphics[scale=0.6]{cadre.pdf}}
\vspace{-370pt}
\hbox{\hskip 250pt\includegraphics[scale=0.4]{clavier.pdf}}
\vspace{-27.5pt}
\hbox{\hskip 15pt\hbox{\hskip 40pt\includegraphics[scale=0.25]{wo0_moi.jpg}}
\hskip-27.5pt\hbox{\hskip 20pt\includegraphics[scale=0.25]{men0_pluriel.jpg}}
}
\vspace{-27pt}
\hbox{\hskip 50pt\tt shi}
\vspace{73pt}
\begin{fig}\label{theshi1}
The panel for {\tt shi} is triggered.
\end{fig}
}

\vtop{
\vspace{-20pt}
\hbox{\hskip 30pt\includegraphics[scale=0.275]{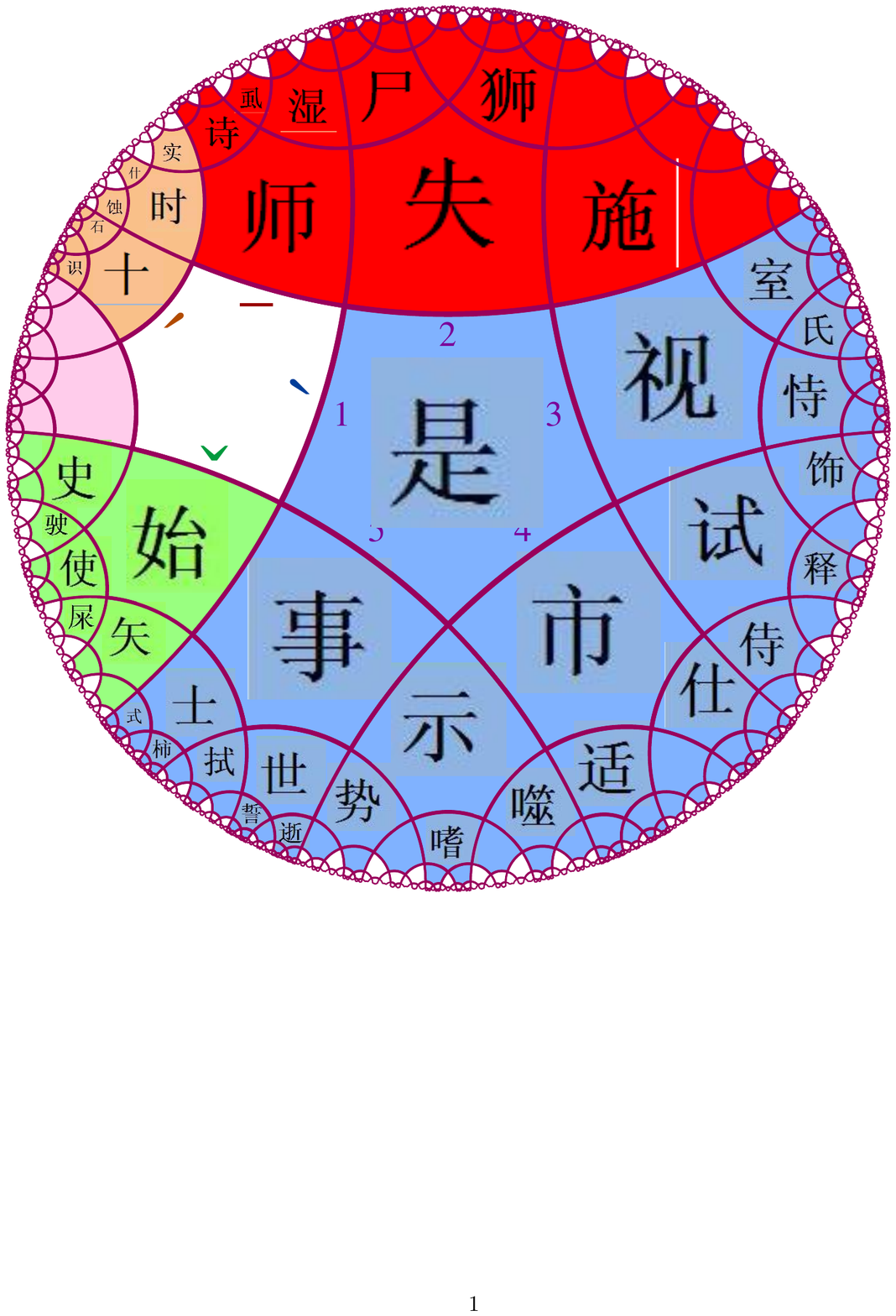}}
\vspace{-465pt}
\hbox{\hskip 30pt\includegraphics[scale=0.6]{cadre.pdf}}
\vspace{-370pt}
\hbox{\hskip 250pt\includegraphics[scale=0.4]{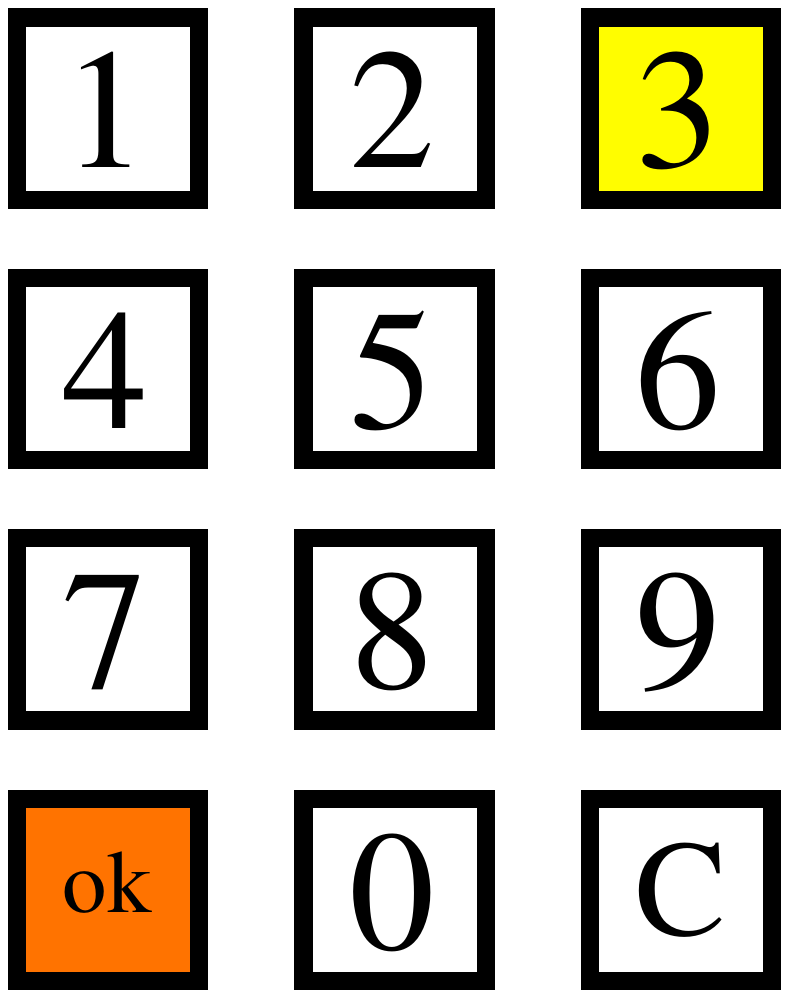}}
\vspace{-27.5pt}
\hbox{\hskip 15pt\hbox{\hskip 40pt\includegraphics[scale=0.25]{wo0_moi.jpg}}
\hskip-27.5pt\hbox{\hskip 20pt\includegraphics[scale=0.25]{men0_pluriel.jpg}}
\hskip-27.5pt\hbox{\hskip 20pt\includegraphics[scale=0.25]{shi0_etre.jpg}}
}
\vspace{53pt}
\begin{fig}\label{theshi2}
The expected character is selected,here \includegraphics[scale=0.20]{shi0_etre.jpg}.
\end{fig}
}

\subsubsection{Additional features}
\label{improve}

    We can now illustrate the other use of the cell phone in case the user does not
clearly understand that {\tt sh} can be accessed without the use of {\tt s} followed 
by~{\tt h}. Figures~\ref{pansshi1}, \ref{pansshi2} and~\ref{pansshi3} show the actions
that the user has to perform if he/she proceeds in this way.

 Note that this problem also arises for~{\tt zh} and~{\tt ch}. A single solution
is possible for the user who would not immediately grasp the phonetic aspect of the proposal.

   The user wishes to first access to~{\tt s}. He/she must first select {\tt z} as shown
in Figure~\ref{pansshi1} by pressing key~7. Then, he/she presses the same key again in order to 
get~{\tt s}, see Figure~\ref{pansshi2}. At last, he/she presses key~0{} in order 
to transform~{\tt s} into~{\tt sh} and also on~{\tt ok} in order to confirm the choice.
Of course, the key~{\tt ok} can be used in a similar way for~{\tt zh} and~{\tt ch}.

We think that after a certain time, the user will understand the logic of the various colours, 
that the consonents are divided into groups. The user will sooner or later realize that it is 
shorter to get~{\tt sh} by selecting first~{\tt zh} and not using the helping key~{\tt 0}. 
He/she will discover this feature when noticing that from any letter on the panel, thanks to
the numbers in the fixed central part, it is possible to go to any other letter, as well 
as to go back to an empty central cell.

\vskip 20pt
\vtop{
\vspace{-150pt}
\hbox{\hskip 50pt\includegraphics[scale=0.35]{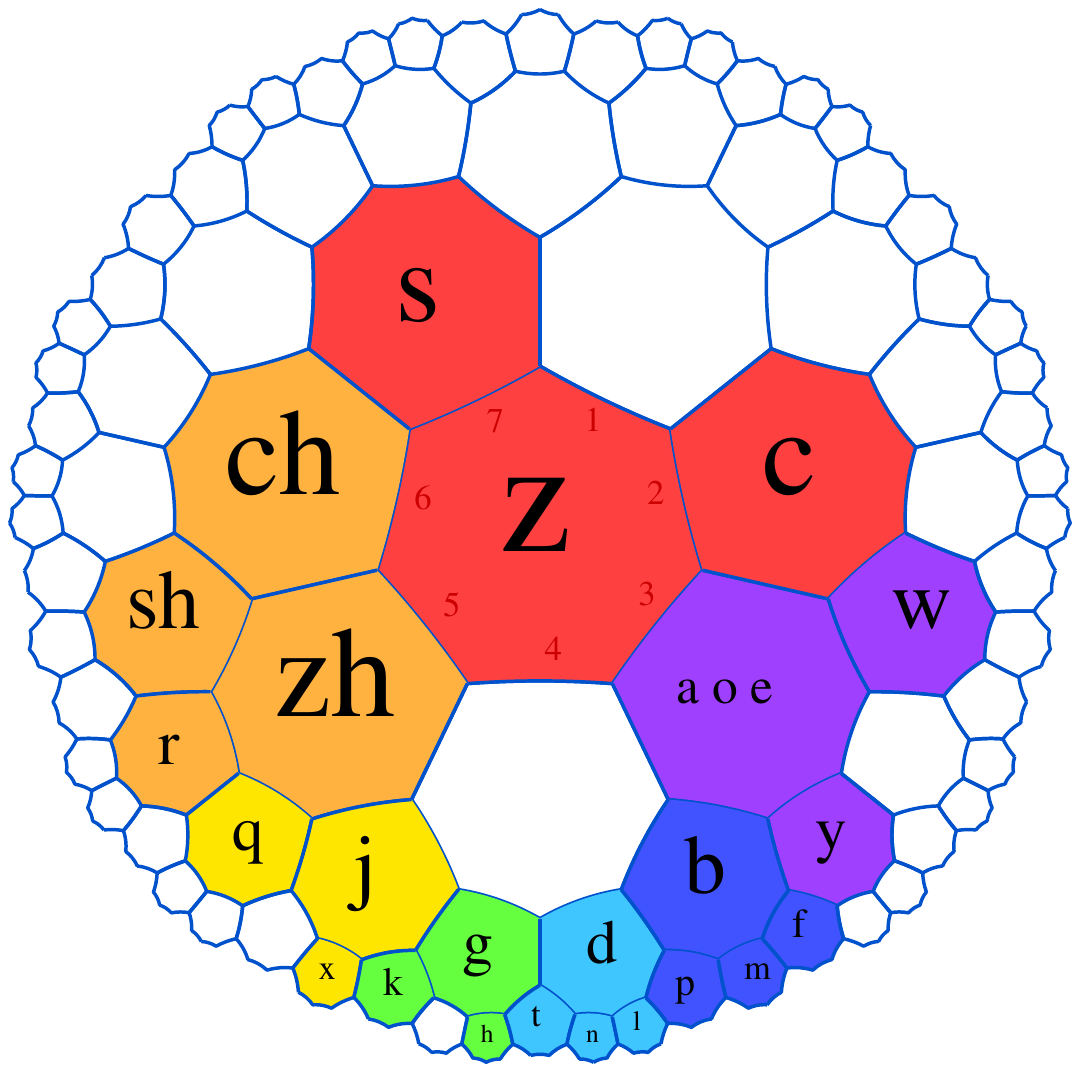}}
\vspace{-390pt}
\hbox{\hskip 30pt\includegraphics[scale=0.6]{cadre.pdf}}
\vspace{-370pt}
\hbox{\hskip 215pt\includegraphics[scale=0.4]{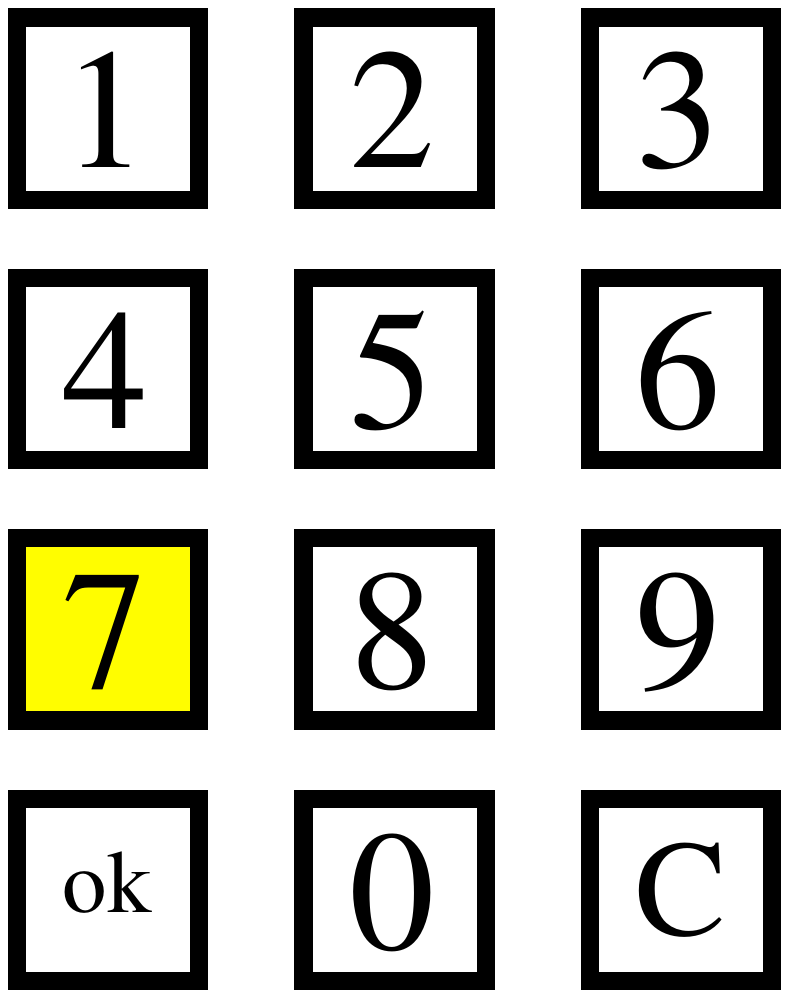}}
\vspace{-27.5pt}
\hbox{\hskip 15pt\hbox{\hskip 40pt\includegraphics[scale=0.25]{wo0_moi.jpg}}
\hskip-27.5pt\hbox{\hskip 20pt\includegraphics[scale=0.25]{men0_pluriel.jpg}}
}
\vspace{55pt}
\begin{fig}\label{pansshi1}
Choosing the letter: accessing {\tt sh} through {\tt s}.
\end{fig}
}

\vtop{
\vspace{-150pt}
\hbox{\hskip 50pt\includegraphics[scale=0.35]{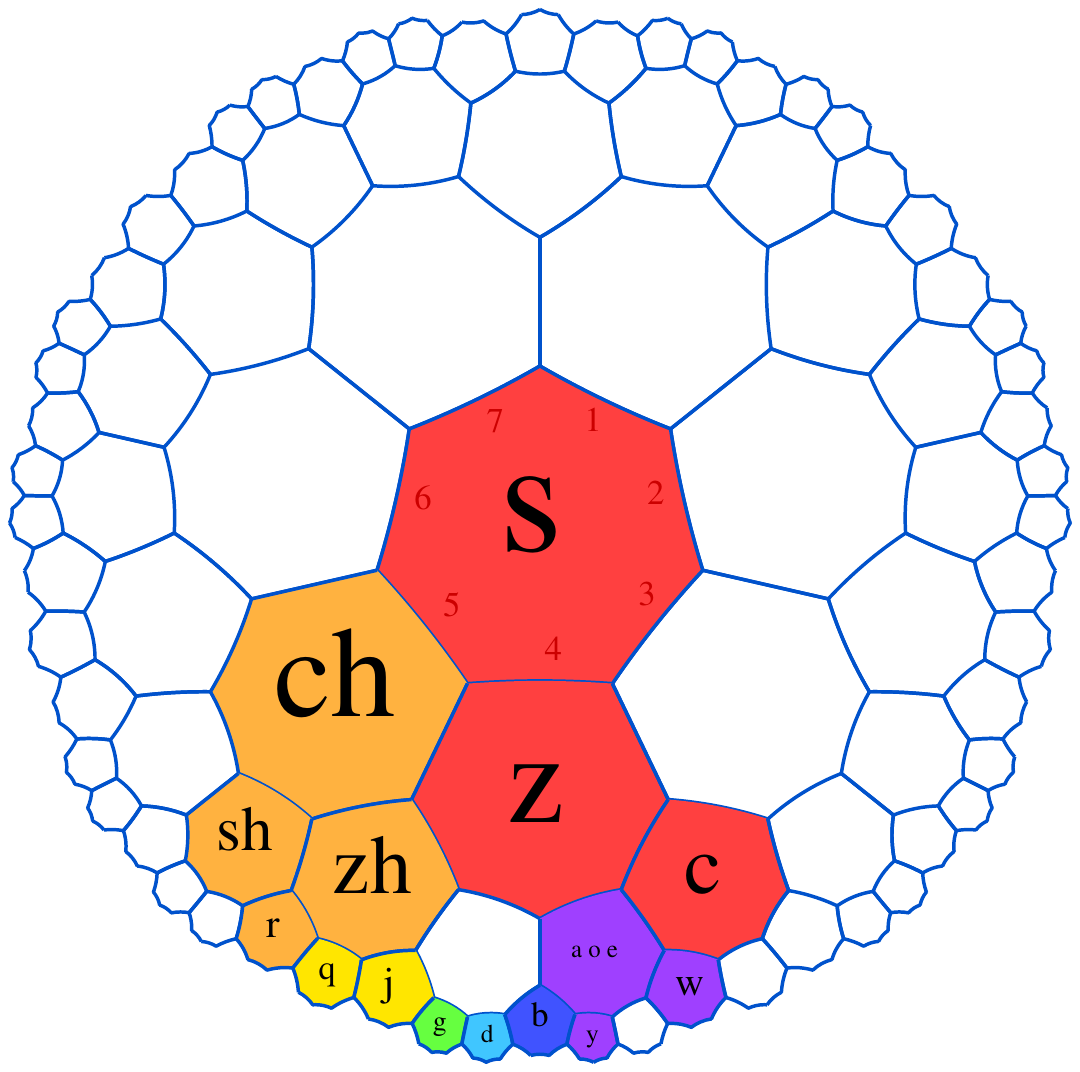}}
\vspace{-390pt}
\hbox{\hskip 30pt\includegraphics[scale=0.6]{cadre.pdf}}
\vspace{-370pt}
\hbox{\hskip 215pt\includegraphics[scale=0.4]{clavier7.pdf}}
\vspace{-27.5pt}
\hbox{\hskip 15pt\hbox{\hskip 40pt\includegraphics[scale=0.25]{wo0_moi.jpg}}
\hskip-27.5pt\hbox{\hskip 20pt\includegraphics[scale=0.25]{men0_pluriel.jpg}}
}
\vspace{-27pt}
\hbox{\hskip 50pt}
\vspace{73pt}
\begin{fig}\label{pansshi2}
The user accessed to {\tt s}.
\end{fig}
}

\vtop{
\vspace{-150pt}
\hbox{\hskip 50pt\includegraphics[scale=0.35]{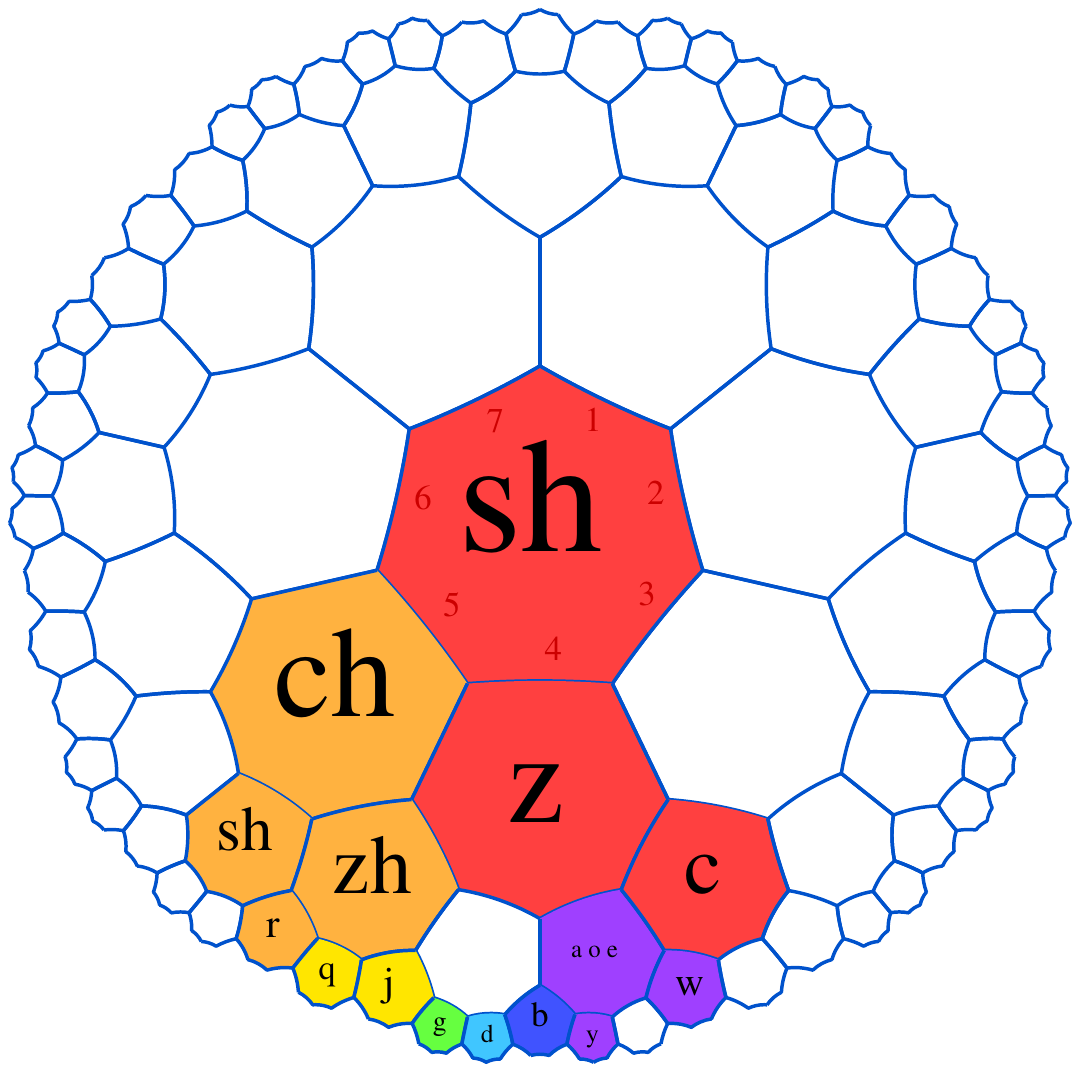}}
\vspace{-390pt}
\hbox{\hskip 30pt\includegraphics[scale=0.6]{cadre.pdf}}
\vspace{-370pt}
\hbox{\hskip 215pt\includegraphics[scale=0.4]{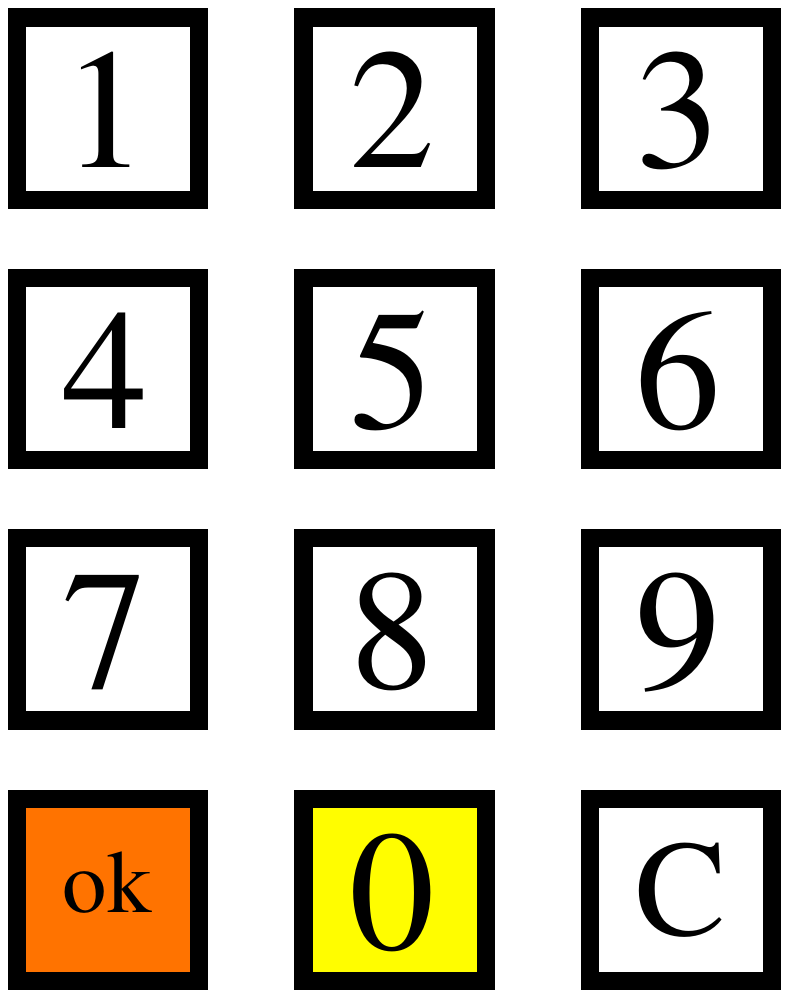}}
\vspace{-27.5pt}
\hbox{\hskip 15pt\hbox{\hskip 40pt\includegraphics[scale=0.25]{wo0_moi.jpg}}
\hskip-27.5pt\hbox{\hskip 20pt\includegraphics[scale=0.25]{men0_pluriel.jpg}}
}
\vspace{-27pt}
\hbox{\hskip 50pt\tt sh}
\vspace{73pt}
\begin{fig}\label{pansshi3}
The user changes {\tt s} to~{\tt sh} and confirms the choice.
\end{fig}
}

   Starting from this point, the vowel panel is triggered and the user performs the actions
already depicted by Figures~\ref{vowshi1}, \ref{vowshi2}, \ref{theshi1} and~\ref{theshi2}.
\vskip 20pt
   Another way to use the cell phone is accessible to the enough trained user. It consists
in numbering the cells used in the panels, mainly for the pin-yin panels. This is illustrated
by Figures~\ref{num_cons} and~\ref{num_vows} for the consonents, for the wowels respectively.

\vtop{
\vspace{-220pt}
\ligne{\hfill\includegraphics[scale=0.475]{bopomofo_7_3.pdf}
\hskip -120pt
\includegraphics[scale=0.475]{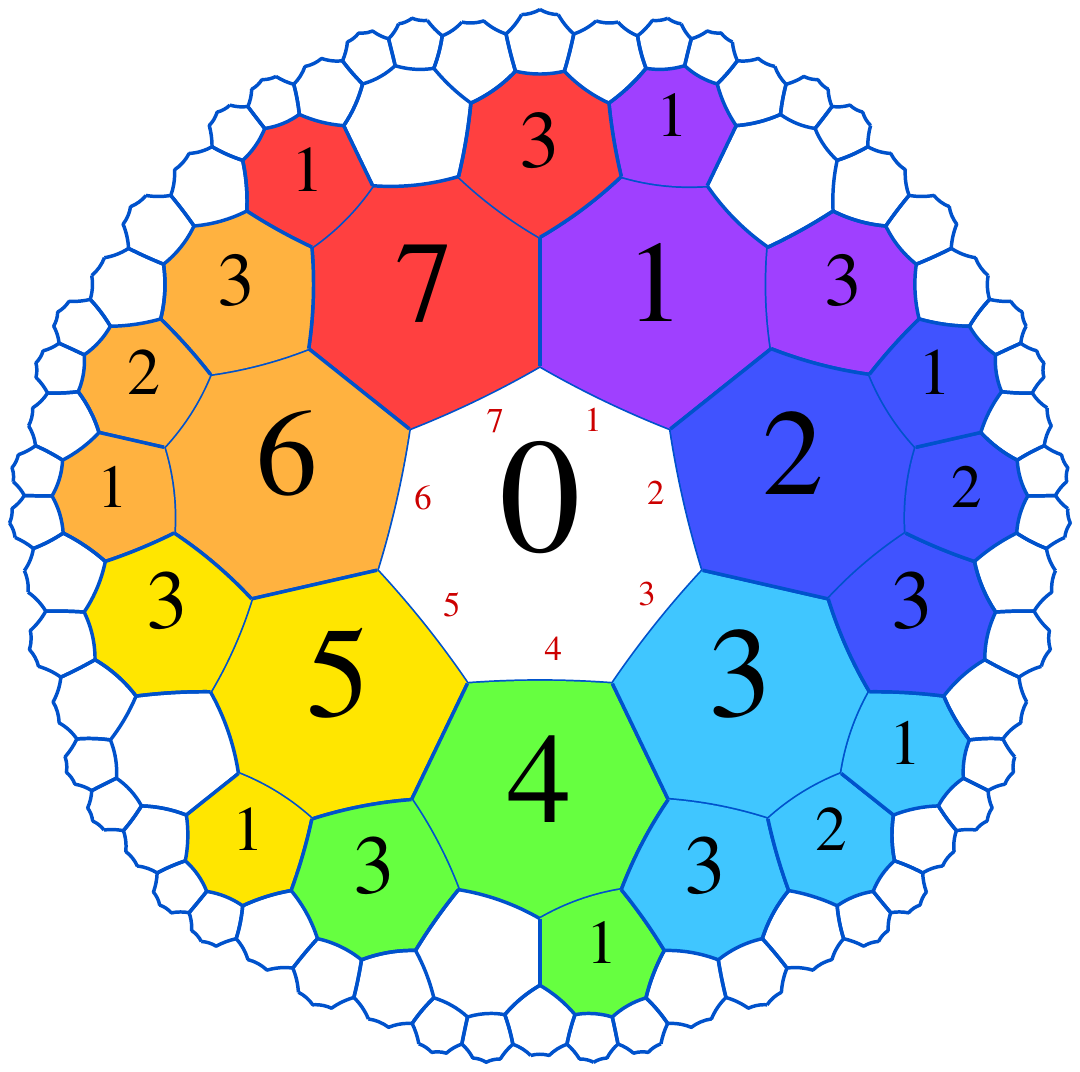}
\hfill}
\vspace{-5pt}
\begin{fig}
\label{num_cons}
\leurre
Numbering the cells on the consonent panel.
\end{fig}
}

   The principle of the numbering is simple. The cells to which the user access by pressing a key
when starting from an empty cell bear the number indicated on the display. As the device expects
a number with two digits, the user dial {\tt 0x}, where {\tt x} is the appropriate number,
or he/she presses {\tt x} and then {\tt ok}. For example, to enter~{\tt j}, the user
forms {\tt 05} or {\tt 5} and then~{\tt ok}. Then, if the letter is a bit further, it is numbered
in its sector, defined by the colour, the numbering being increasing on each ring around the
central cell, starting from the closest to the center and clock-wise turning around the central
cell. We can see the application of this rule on both Figures~\ref{num_cons} and~\ref{num_vows}.

\vtop{
\vspace{-220pt}
\ligne{\hfill\includegraphics[scale=0.475]{voyelles_7_3.pdf}
\hskip -120pt
\includegraphics[scale=0.475]{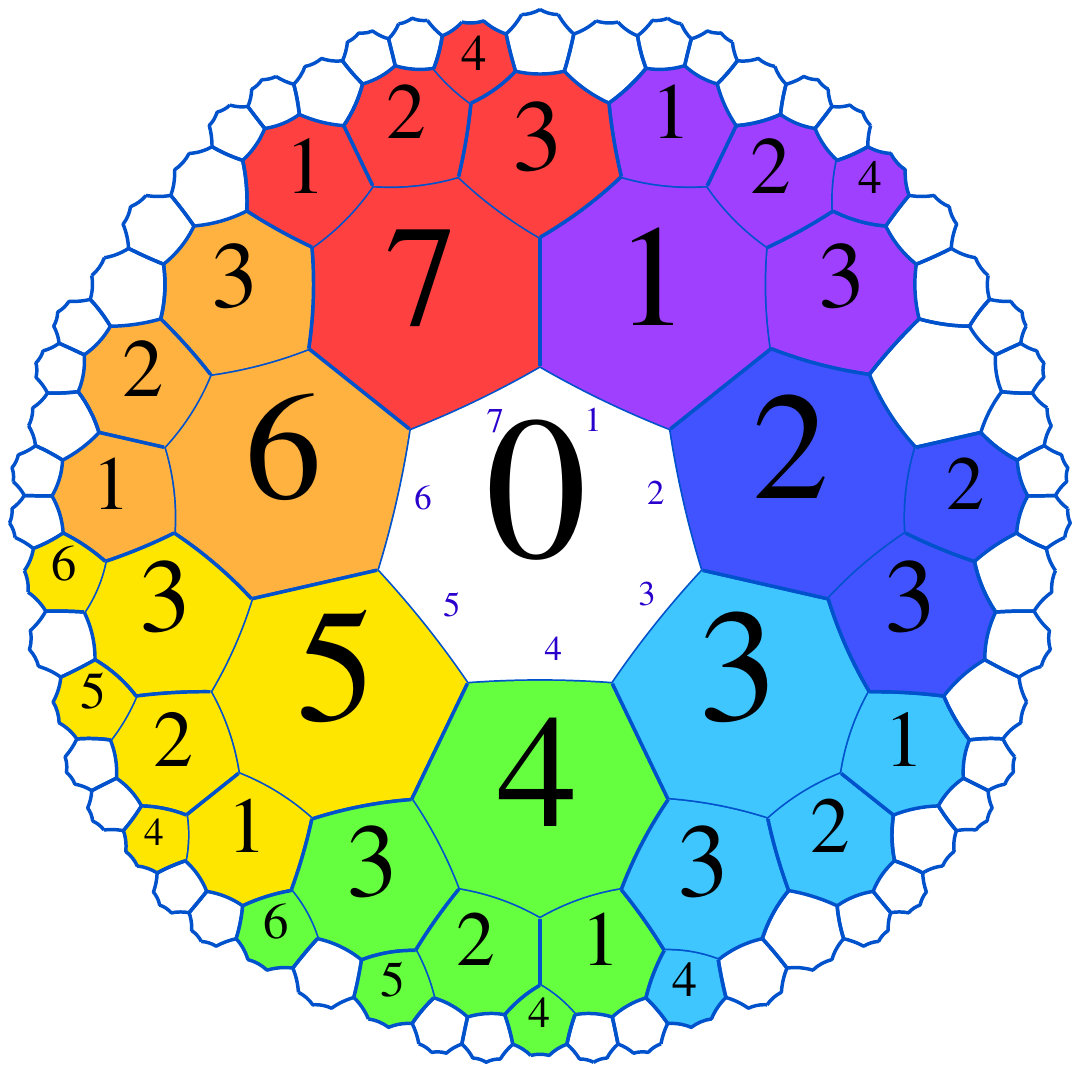}
\hfill}
\vspace{-5pt}
\begin{fig}
\label{num_vows}
\leurre
Numbering the cells on the vowel panel.
\end{fig}
}

   As an example, {\tt sh} is numbered {\tt 62} in the consonent panel: the user 
presses~{\tt 6} and then~{\tt 2}. This convention requires the minimal memory effort for 
the user. As another example, this time in the vowel panel, {\tt uan} is 
numbered~{\tt 55} and is accessed by clicking {\tt 5} and then~{\tt 5} again. For both
panels, the numbering of the central cell reminds the numbers of the sectors and the 
orientation of the numbering in the other further rings. 

    However, it should be noted that this numbering is no more possible for the panels of
characters, especially when there are many of them. There is a possible numbering of the 
tiles on which the coordinate system evoked in Subsection~\ref{hyptilings} is based, but this 
is out of scope for most users. The user will certainly shortly realize that counting the 
tiles to find out a number is much longer than clicking on two or three keys in order to 
access the appropriate character.

\subsection{The example on a tactile screen device}
\label{tactile}

    Let us know look at the same examples in the case of a tactile screen device. 
Figure~\ref{tact_w0} shows the first screen seen by the user: the consonent panel.

\vtop{
\vspace{-150pt}
\hbox{\hskip 60pt\includegraphics[scale=0.35]{bopomofo_7_3.pdf}}
\vspace{-350pt}
\hbox{\hskip 30pt\includegraphics[scale=0.5]{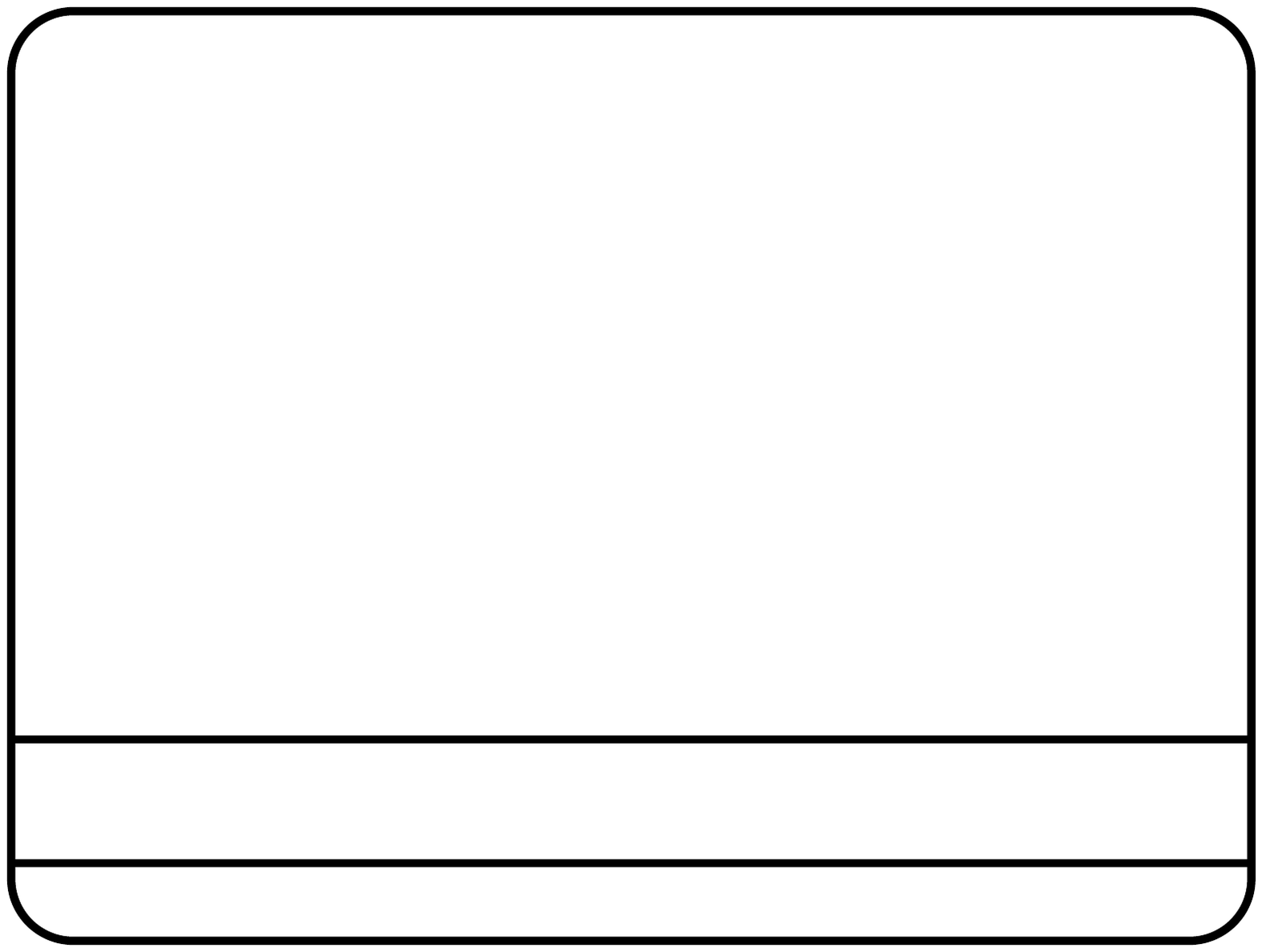}}
\begin{fig}\label{tact_w0}
The consonent panel on a tactile screen.
\end{fig}
}

Then, the user presses the screen on the spot where the desired consonent is, see 
Figure~\ref{tact_w1}. Here, the consonent is~{\tt w}.

\vtop{
\vspace{-135pt}
\hbox{\hskip 60pt\includegraphics[scale=0.35]{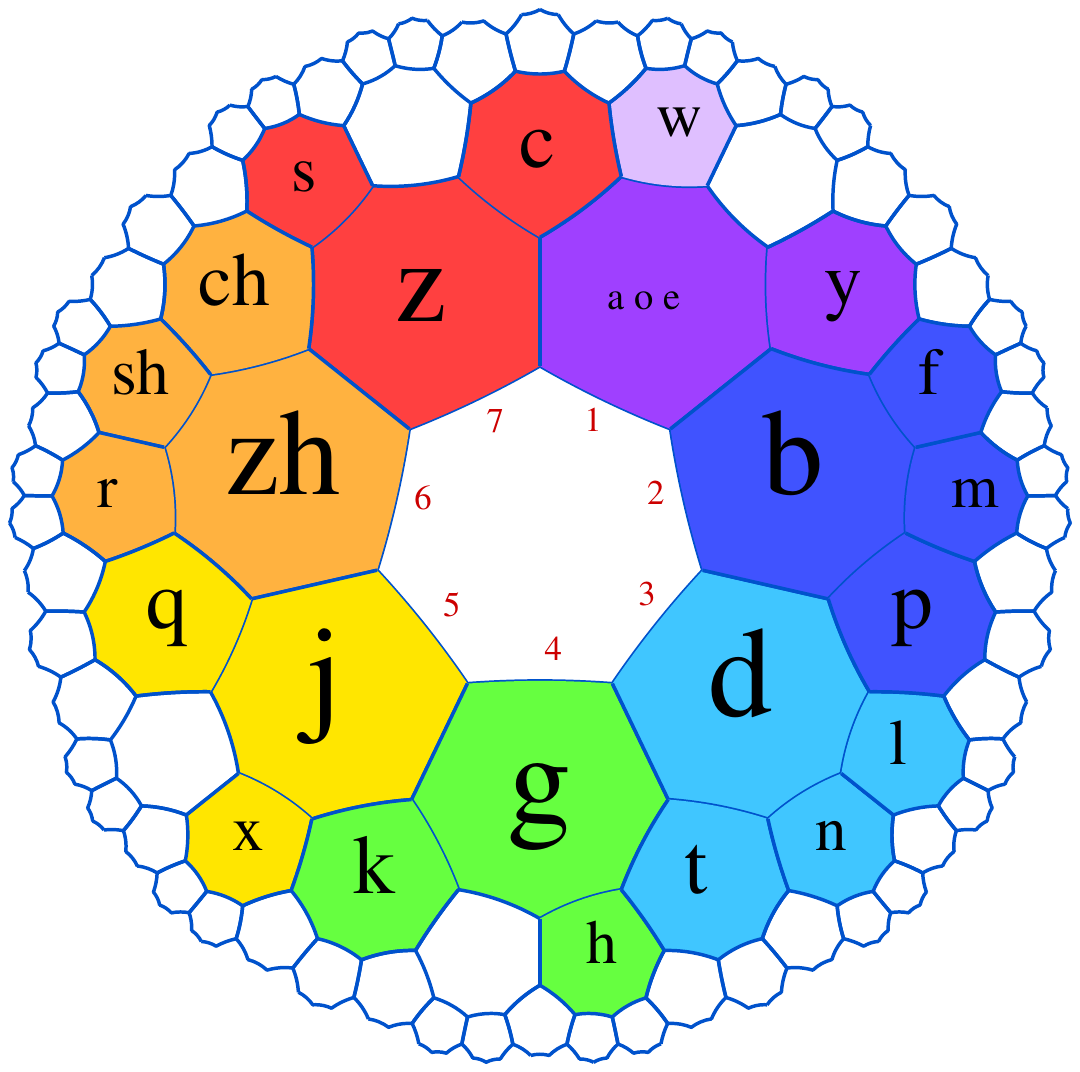}}
\vspace{-350pt}
\hbox{\hskip 30pt\includegraphics[scale=0.5]{cadre2.pdf}}
\vspace{-47.5pt}
\hbox{\hskip 50pt\tt w}
\vspace{32.5pt}
\begin{fig}\label{tact_w1}
The consonent panel on a tactile screen: the user choosed~{\tt w}.
\end{fig}
\vskip 15pt
}

Then, the vowel panel appears, as shown by~Figure~\ref{tact_o0}. The user can now 
choose the desired vowel: it is enough to press on the spot of the screen where the
corresponding tile is.

\vtop{
\vspace{-115pt}
\hbox{\hskip 60pt\includegraphics[scale=0.35]{voyelles_7_3.pdf}}
\vspace{-350pt}
\hbox{\hskip 30pt\includegraphics[scale=0.5]{cadre2.pdf}}
\vspace{-47.5pt}
\hbox{\hskip 50pt\tt w}
\vspace{32.5pt}
\begin{fig}\label{tact_o0}
The vowel panel appears.
\end{fig}
\vskip 15pt
}

   The user chosed the appropriate vowel as can be seen on Figure~\ref{tact_o1}.
This triggers the panel of characters corresponding to the pin-yin~{\tt wo}, see
Figure~\ref{test_wo0}.

\vtop{
\vspace{-130pt}
\hbox{\hskip 60pt\includegraphics[scale=0.35]{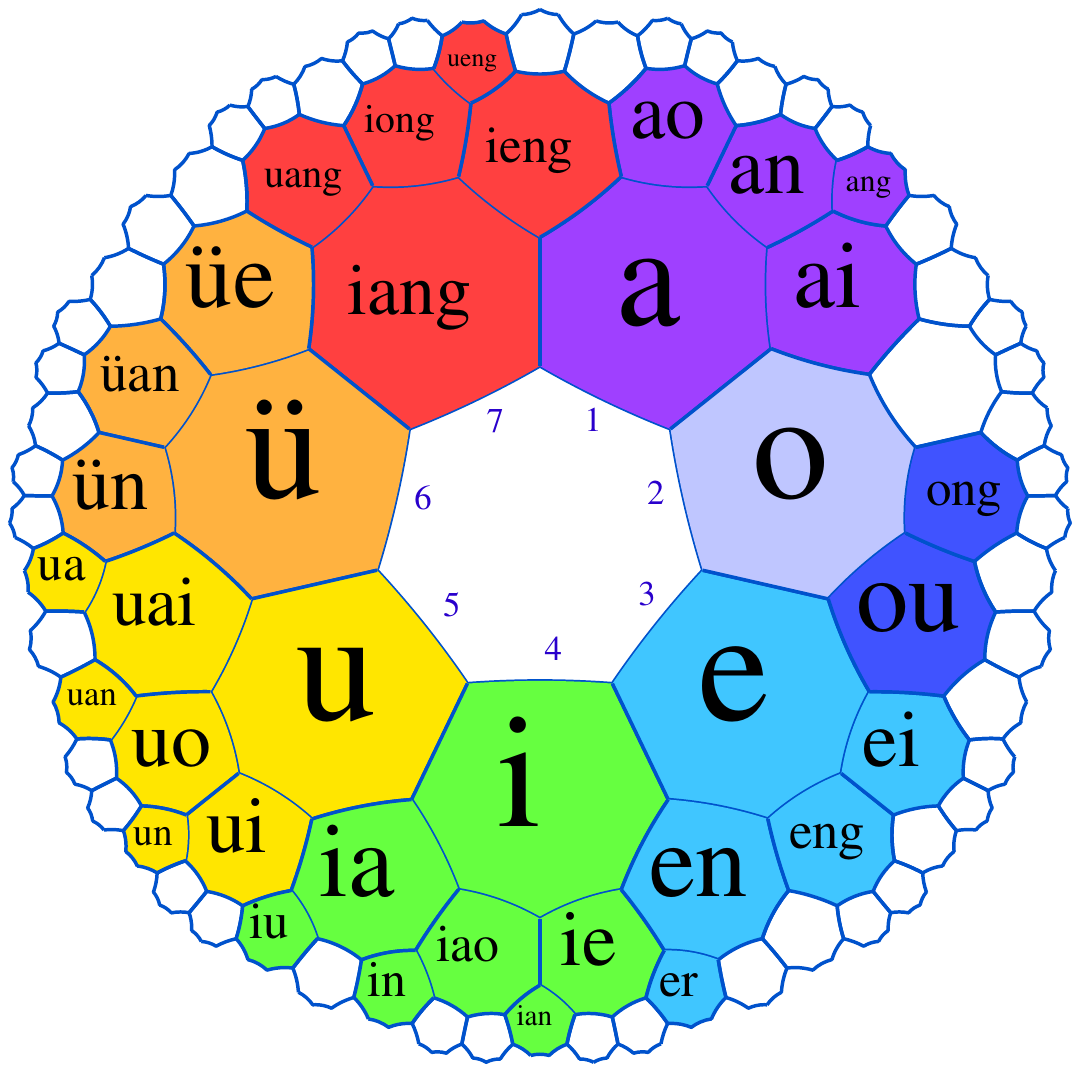}}
\vspace{-350pt}
\hbox{\hskip 30pt\includegraphics[scale=0.5]{cadre2.pdf}}
\vspace{-47.5pt}
\hbox{\hskip 50pt\tt wo}
\vspace{32.5pt}
\begin{fig}\label{tact_o1}
Then, the vowel panel: the user choosed~{\tt o}: {\tt wo} is formed.
\end{fig}
}

\vtop{
\vspace{-100pt}
\hbox{\hskip 40pt\includegraphics[scale=0.275]{test_wo.pdf}}
\vspace{-420pt}
\hbox{\hskip 30pt\includegraphics[scale=0.5]{cadre2.pdf}}
\vspace{-47.5pt}
\hbox{\hskip 50pt\tt wo}
\vspace{32.5pt}
\begin{fig}\label{test_wo0}
Then, the characters for the pin-yin {\tt wo} are displayed.
\end{fig}
}

   The user chooses the needed character as shown in Figure~\ref{test_wo1}. Once chosen,
the character appears on the control screen: it is here too placed under the
screen for the consonent and vowel panels.

\vtop{
\hbox{\hskip 40pt\includegraphics[scale=0.275]{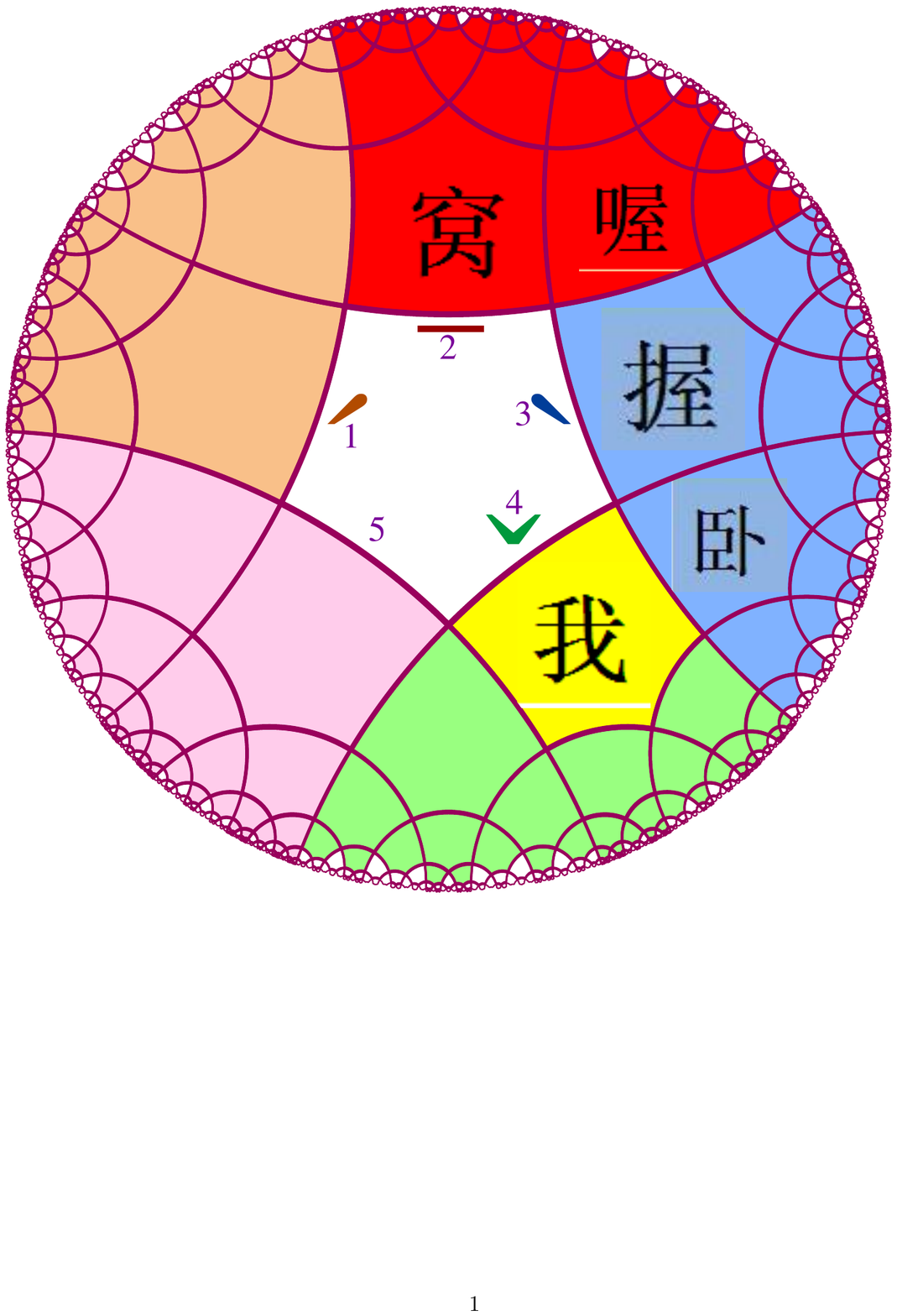}}
\vspace{-420pt}
\hbox{\hskip 30pt\includegraphics[scale=0.5]{cadre2.pdf}}
\vspace{-40pt}
\hbox{\hskip 25pt\hbox{\hskip 40pt\includegraphics[scale=0.225]{wo0_moi.jpg}}
}
\vspace{20pt}
\begin{fig}\label{test_wo1}
The user choosed the appropriate character: \includegraphics[scale=0.20]{wo0_moi.jpg}.
\end{fig}
}

We have exactly the same situation for the choice of~{\tt shi}. Figures~\ref{pan_shi0}
and~\ref{pan_shi1} illustrate the selection of~{\tt sh}, which requires a single
gesture and which does not raise the problem of conflict between the spelling and
the phonetics. Then, Figures~\ref{vow_shi0} and~\ref{vow_shi1} illustrate the choice
of the vowel~{\tt i}. This triggers the panel of characters associated with~{\tt shi}
see Figure~\ref{test_shi0}. A single gesture of the user allows to select
the needed character, see Figure~\ref{test_shi1}.

\vtop{
\vspace{-150pt}
\hbox{\hskip 60pt\includegraphics[scale=0.35]{bopomofo_7_3.pdf}}
\vspace{-350pt}
\hbox{\hskip 30pt\includegraphics[scale=0.5]{cadre2.pdf}}
\vspace{-40pt}
\hbox{\hskip 25pt\hbox{\hskip 40pt\includegraphics[scale=0.225]{wo0_moi.jpg}}
\hskip-45pt\hbox{\hskip 40pt\includegraphics[scale=0.225]{men0_pluriel.jpg}}
}
\vspace{20pt}
\begin{fig}\label{pan_shi0}
Again the consonent panel.
\end{fig}
}

\vtop{
\vspace{-150pt}
\hbox{\hskip 60pt\includegraphics[scale=0.35]{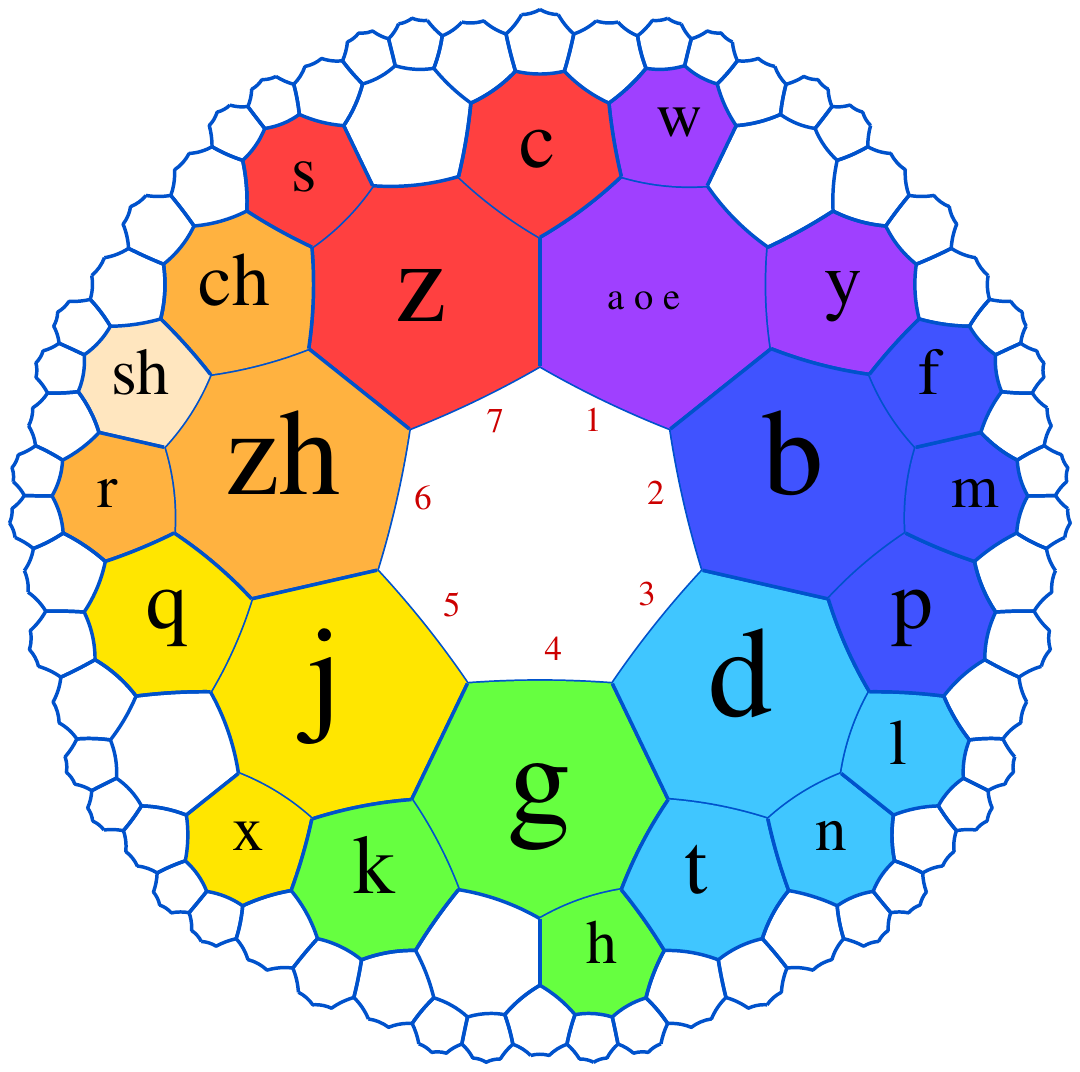}}
\vspace{-350pt}
\hbox{\hskip 30pt\includegraphics[scale=0.5]{cadre2.pdf}}
\vspace{-47.5pt}
\hbox{\hskip 50pt\tt sh}
\vspace{-5pt}
\hbox{\hskip 25pt\hbox{\hskip 40pt\includegraphics[scale=0.225]{wo0_moi.jpg}}
\hskip-45pt\hbox{\hskip 40pt\includegraphics[scale=0.225]{men0_pluriel.jpg}}
}
\vspace{20pt}
\begin{fig}\label{pan_shi1}
The direct selection of~{\tt sh}.
\end{fig}
}

\vtop{
\vspace{-150pt}
\hbox{\hskip 60pt\includegraphics[scale=0.35]{voyelles_7_3.pdf}}
\vspace{-350pt}
\hbox{\hskip 30pt\includegraphics[scale=0.5]{cadre2.pdf}}
\vspace{-47.5pt}
\hbox{\hskip 50pt\tt sh}
\vspace{-5pt}
\hbox{\hskip 25pt\hbox{\hskip 40pt\includegraphics[scale=0.225]{wo0_moi.jpg}}
\hskip-45pt\hbox{\hskip 40pt\includegraphics[scale=0.225]{men0_pluriel.jpg}}
}
\vspace{20pt}
\begin{fig}\label{vow_shi0}
Now, the vowel panel.
\end{fig}
}

\vtop{
\vspace{-150pt}
\hbox{\hskip 60pt\includegraphics[scale=0.35]{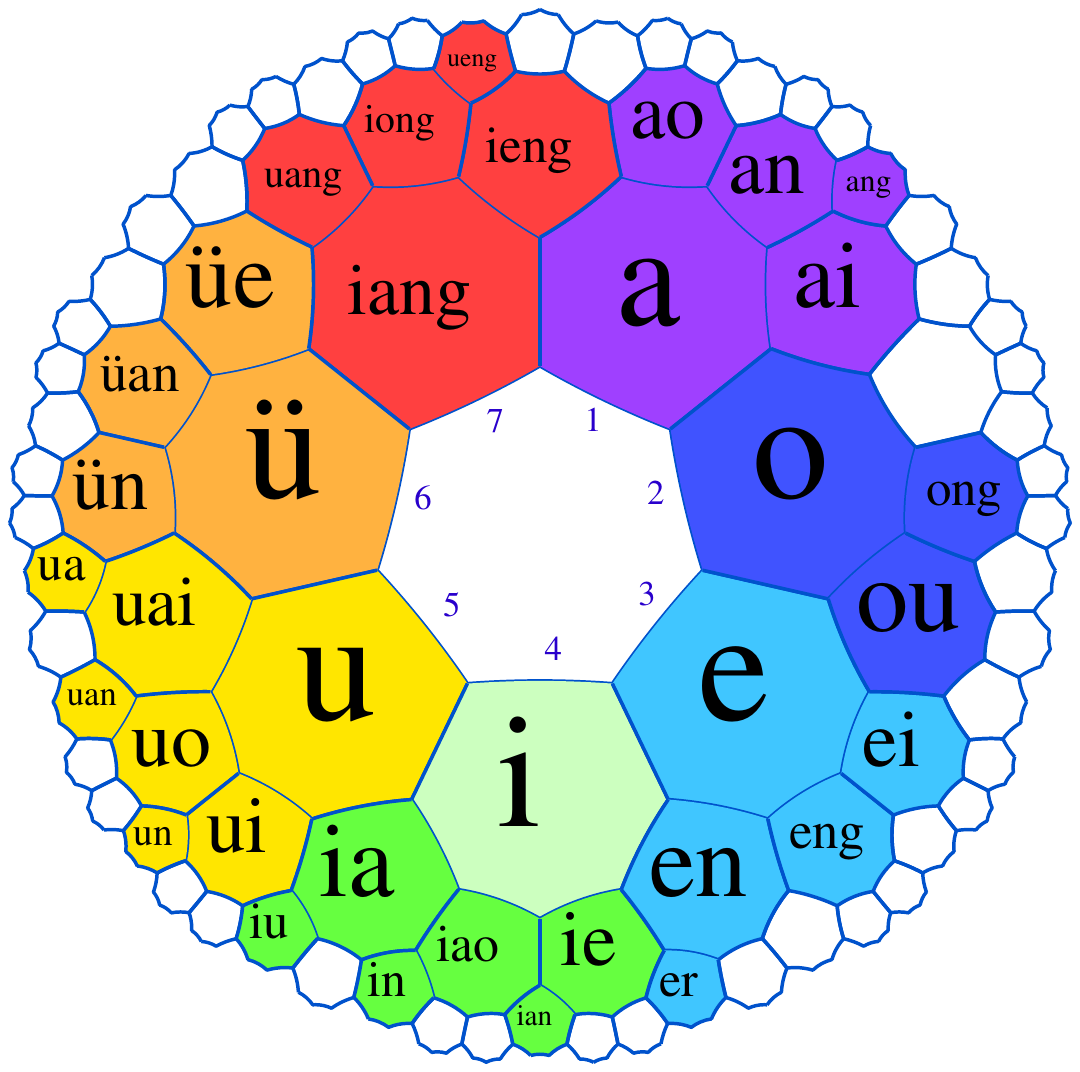}}
\vspace{-350pt}
\hbox{\hskip 30pt\includegraphics[scale=0.5]{cadre2.pdf}}
\vspace{-47.5pt}
\hbox{\hskip 50pt\tt shi}
\vspace{-5pt}
\hbox{\hskip 25pt\hbox{\hskip 40pt\includegraphics[scale=0.225]{wo0_moi.jpg}}
\hskip-45pt\hbox{\hskip 40pt\includegraphics[scale=0.225]{men0_pluriel.jpg}}
}
\vspace{20pt}
\begin{fig}\label{vow_shi1}
At present, {\tt shi} is selected.
\end{fig}
}

\vtop{
\hbox{\hskip 40pt\includegraphics[scale=0.275]{test_shi.pdf}}
\vspace{-420pt}
\hbox{\hskip 30pt\includegraphics[scale=0.5]{cadre2.pdf}}
\vspace{-47.5pt}
\hbox{\hskip 50pt\tt shi}
\vspace{-5pt}
\hbox{\hskip 25pt\hbox{\hskip 40pt\includegraphics[scale=0.225]{wo0_moi.jpg}}
\hskip-45pt\hbox{\hskip 40pt\includegraphics[scale=0.225]{men0_pluriel.jpg}}
}
\vspace{20pt}
\begin{fig}\label{test_shi0}
The characters for~{\tt shi} are displayed.
\end{fig}
}

\vtop{
\hbox{\hskip 40pt\includegraphics[scale=0.275]{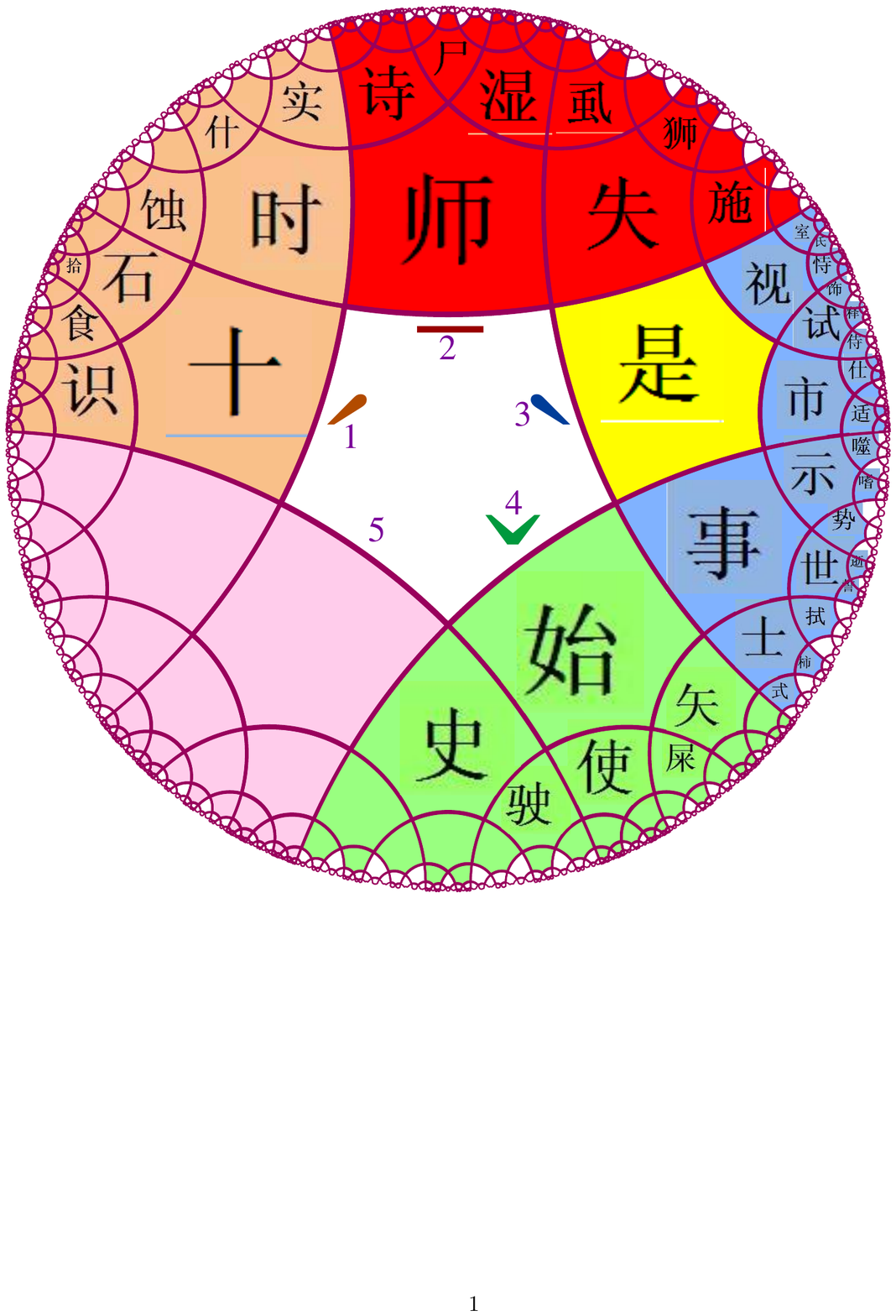}}
\vspace{-420pt}
\hbox{\hskip 30pt\includegraphics[scale=0.5]{cadre2.pdf}}
\vspace{-40pt}
\hbox{\hskip 25pt\hbox{\hskip 40pt\includegraphics[scale=0.225]{wo0_moi.jpg}}
\hskip-45pt\hbox{\hskip 40pt\includegraphics[scale=0.225]{men0_pluriel.jpg}}
\hskip-45pt\hbox{\hskip 40pt\includegraphics[scale=0.225]{shi0_etre.jpg}}
}
\vspace{20pt}
\begin{fig}\label{test_shi1}
The user can now select the appropriate character for {\tt shi}, here  
\includegraphics[scale=0.20]{shi0_etre.jpg}.
\end{fig}
}

  On tablets, where the screen is bigger, we can afford an interesting improvement in order
to raise the efficiency of the method. The selection of the pin-yin can be facilitated
by presenting to the user both the panel of consonents and the panel of vowels at the
same time, as shown in Figure~\ref{bipy_shi0}.

\vtop{
\vspace{-150pt}
\hbox{\hskip 50pt\includegraphics[scale=0.325]{bopomofo_7_3.pdf}
\hskip-100pt\includegraphics[scale=0.325]{voyelles_7_3.pdf}}
\vspace{-345pt}
\hbox{\hskip 30pt\includegraphics[scale=0.5]{cadre2.pdf}}
\begin{fig}\label{bipy_shi0}
Both panels, for the consonents and for the vowels, are displayed at the same time.
\end{fig}
}

Figure~\ref{bipy_shi1} shows that the user can select both the consonent and the
vowel on the same screen, which is somehow faster.

\vtop{
\vspace{-125pt}
\hbox{\hskip 50pt\includegraphics[scale=0.325]{bopo_sh.pdf}
\hskip-100pt\includegraphics[scale=0.325]{voye_i.pdf}}
\vspace{-345pt}
\hbox{\hskip 30pt\includegraphics[scale=0.5]{cadre2.pdf}}
\vspace{-47.5pt}
\hbox{\hskip 50pt\tt shi}
\vspace{35pt}
\begin{fig}\label{bipy_shi1}
The user can now select the appropriate pin-yin, whatever the order of the
selection, here {\tt shi}.
\end{fig}
}

   After this selection, the screen displays the panel of the characters for~{\tt shi}
as illustrated by Figure~\ref{test_shi0}.

\subsection{The predictive typing}
\label{predictive}

   We turn now to the predictive typing we mentioned in Section~\ref{main}. We indicated there
that many systems for typing texts in Chinese language make use of such a facility. We can
do the same here. Such a system is based on a large data set where connections between
words are registered. 

   We may imagine that each letter entered in the system raises a set of
possible words which are ordered by the system according to some algorithm. The user
confirms the first choice or selects one of the proposed choices or goes back to
a previous situation if the proposed selections do not fit what he/she has in mind. Accordingly,
the behaviour of the system consists in defining paths in trees rooted at the intial entry
after the last past session. 

    We give the user the possibilitly to switch on the predictive
typing or to switch it off. By default, the predictive system is switched off. 

If the system is switched on, two keys or buttons are at the disposal of the user.
The first one makes a selection between two degrees of predictivity: in the lowest degree
the system suggests choices a characters for the pin-yin while the user is dealing
with the panel of consonents. These choices are displayed on the control screen. 
Then, the selection is performed by clicking
on a number which is the order of the desired character in the list suggested by
the system or, on a device with a tactile screen, by pressing on the character.
In the highest degree, to the just described facility, the system offers an additional
one. Once a character is selected, the system suggests possible characters which may follow 
what is already recorder on the linear display. As an example, after
\includegraphics[scale=0.20]{wo0_moi.jpg}{}
is entered, the system may display:
\includegraphics[scale=0.20]{men0_pluriel.jpg}, 
\includegraphics[scale=0.20]{hao0_bon.jpg}{},
\includegraphics[scale=0.20]{shi0_etre.jpg},
\includegraphics[scale=0.28]{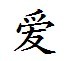}, 
\includegraphics[scale=0.28]{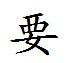}, 
\includegraphics[scale=0.28]{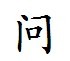},
\includegraphics[scale=0.28]{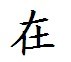}
 and a lot of other possibilities.

\vtop{
\vspace{-150pt}
\hbox{\hskip 50pt\includegraphics[scale=0.35]{bopomofo_7_3.pdf}}
\vspace{-390pt}
\hbox{\hskip 30pt\includegraphics[scale=0.6]{cadre.pdf}}
\vspace{-370pt}
\hbox{\hskip 215pt\includegraphics[scale=0.4]{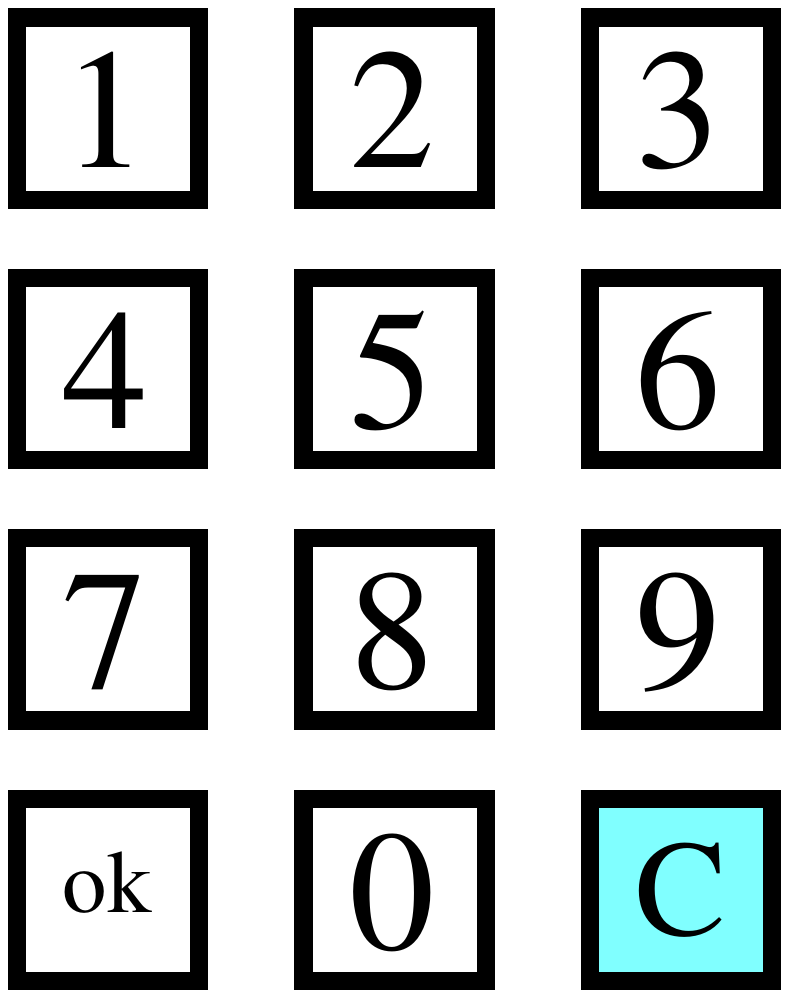}}
\vspace{45pt}
\begin{fig}\label{panpred}
The key for controlling the predictive system.
\end{fig}
}

   In both degrees, a key or a button allows him/her
to go back to a previous step in the currently explored path if needed. If the key or the
button is pressed, the consonent and vowel panels are again raised for determining the
new pin-yin or, depending on what was the last step in the path, the panel of characters for
a given pin-yin is again displayed in order the reader can select a character different from
the one suggested by the system. Figure~\ref{panpred} illustrates the situation for
a cell phone.

Figure~\ref{tabpred} does the same for a smartphone, an Ipad or a tablet: a button is offered
the user in order to go backwards if needed, as long as wished.

\vtop{
\vspace{-130pt}
\hbox{\hskip 50pt\includegraphics[scale=0.35]{bopomofo_7_3.pdf}}
\vspace{-345pt}
\hbox{\hskip 30pt\includegraphics[scale=0.5]{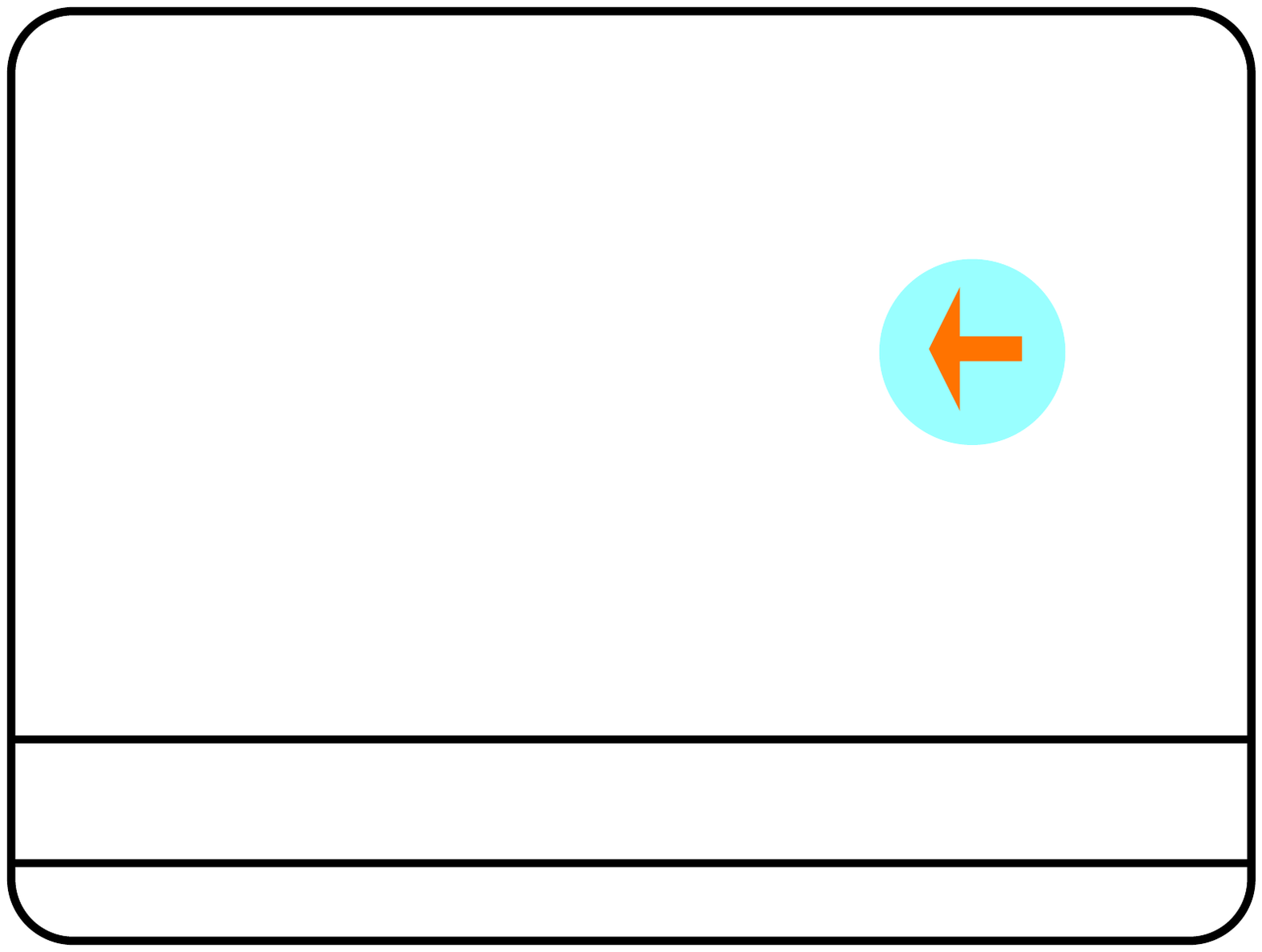}}
\begin{fig}\label{tabpred}
The button for controlling the predictive system.
\end{fig}
}

   Another important point connected to the predictive system is the choice performed by the
system. Usually, the system takes into account the statistical frequency of the characters
as established by many sources. It may also take into account the personal frequency by the
user. As an example, if the user has a close friend whose name has a pin-yin associated with very 
frequent characters, the friend's name will occur sufficiently enough to change the statistics
of this character. Our software will take this into account and it will give priority to
the user personal statistics. The software counts each character used by the user and 
each character receives a weight which is the number of its use by the user. The heighest 
the score, the closer the place of the character to the central cell in the panel associated
to the corresponding pin-yin. The software will keep these statistics from one session to
the other and, during a session, it will permanently update it according to what 
the user did.

\subsection*{Acknowledgment}

   The second author has completed her sudies at Jianghan University, Wuhan, which helped her
so much.


\begin{thebibliography}{5}
\bibitem{mmbook1}
M. Margenstern,
Cellular Automata in Hyperbolic Spaces, Volume 1, Theory,
{\it OCP}, Philadelphia, (2007), 422p.

\bibitem{mmbook2}
M. Margenstern,
Cellular Automata in Hyperbolic Spaces, Volume 2, Computations and Implementations, 
{\it OCP}, Philadelphia, (2008), 360p.

\bibitem{mmbook3}
M. Margenstern,
Small Universal Cellular Automata in Hyperbolic Spaces, A Collection of Jewels,
{\it Springer}, (2013), 320p.

\bibitem{acri}
M. Margenstern, B. Martin, H. Umeo, S. Yamano, K. Nishioka,
A Proposal for a Japanese Keyboard on Cellular Phones,
{\it Lecture Notes in Computer Science}, {\bf 5191}, (2008), 299-306, Proceedings of
{\bf ACRI'2008}, Yokohama, Japan, Sept. 24-26, 2008.


\end{thebibliography}
\end{document}